\pdfoutput=1
\documentclass{aastex}
\usepackage{hyperref,amsmath}
\usepackage{threeparttable}

\newcommand{\galfit}{G\sc alfit\rm}
 
\newcommand{\Msun}{M$_{\odot}$}
\newcommand{\Mpyr}{M$_\odot$ yr$^{-1}$}
\newcommand{\Lsun}{L$_\odot$}

\title{\large A magnified view of star formation at $z$=0.9 from two lensed galaxies\footnote{Based on new and archival observations made with the NASA/ESA Hubble Space Telescope, obtained at the Space Telescope Science Institute, which is operated by the Association of Universities for Research in Astronomy, Inc., under NASA contract NAS 5-26555. New observations are associated with program \# 11678.} \footnote{\emph{Herschel} is an ESA space observatory with science instruments provided by European-led Principal Investigator consortia and with important participation from NASA.}}
\author{Alice Olmstead\altaffilmark{1}, Jane R. Rigby\altaffilmark{2}, Mark Swinbank\altaffilmark{3}, Sylvain Veilleux\altaffilmark{1,4,5}}
\altaffiltext{1}{Department of Astronomy, University of Maryland, College Park, MD 20742, USA}
\altaffiltext{2}{Observational Cosmology Lab, NASA Goddard Space Flight Center, Greenbelt, MD 20771, USA}
\altaffiltext{3}{Institute for Computational Cosmology, Department of Physics, Durham University, South Road, Durham DH1 3LE}
\altaffiltext{4}{Astroparticle Physics Laboratory, NASA Goddard Space Flight Center, Greenbelt, MD 20771, USA}
\altaffiltext{5}{Max-Planck-Institut f\"ur extraterrestrische Physik, Postfach 1312, D-85741 Garching, Germany}

\received{2013 February 26}
\accepted{2014 June 25}
\slugcomment{Published as \href {``http://iopscience.iop.org/1538-3881/148/4/65/''}{2014, The Astronomical Journal, 148, 65}}

\begin{abstract}
We present new narrow-band H$\alpha$ imaging from the \textit{Hubble Space Telescope} of two $z$=0.91 galaxies that have been lensed by foreground galaxy cluster Abell 2390. These data probe spatial scales as small as $\sim$0.3 kpc, providing a magnified look at the morphology of star formation at an epoch when the global star formation rate was high.  However, dust attenuates our spatially resolved star formation rate (SFR) indicators, the H$\alpha$ and rest-UV emission, and we lack a direct measurement of extinction.  Other studies have found
that ionized gas in galaxies tends to be roughly 50\% more obscured than stars; however, given an unextincted measurement of the SFR we can quantify the relative stellar to nebular extinction and the extinction in H$\alpha$.  We infer SFRs from \emph{Spitzer} and \emph{Herschel} mid- to far-infrared observations and compare these to integrated H$\alpha$ and rest-UV SFRs; this yields stellar to nebular extinction ratios consistent with previous studies.  We take advantage of high spatial resolution and contextualize these results in terms of the source-plane morphologies, comparing the distribution of H$\alpha$ to that of the rest-frame UV and optical light.  In one galaxy, we measure separate SFRs in visually distinct clumps, but can set only a lower limit on the extinction and thus the star formation. Consequently, the data are also consistent with there being an equal amount of extinction along the lines of sight to the ionized gas as to the stars..  Future observations in the far-infrared could settle this by mapping out the dust directly.
\end{abstract}
\begin{document}
\maketitle
\section{Introduction}

Over the past two decades, significant progress has been made to understand the global star formation history of the Universe.  The global star formation rate (SFR) density was much higher in the past:  it plateaued between $z$=4--1 at a rate $\gtrsim10\times$ higher than the rate at $z$=0 (e.g.,\ reviews in  \citealt{Hopkins06}, \citealt{Bouwens10}, and \citealt{Shapley11}.)    What is measured for these distant galaxies are generally integrated quantities, for example the integrated rest-frame ultraviolet spectral energy distribution (SED), from which the star formation rate can be inferred.  Galaxies at $z\ga$1 are generally not spatially well resolved: 0.1\arcsec\ (the typical limit of ground or space-based studies) corresponds to 0.8~kpc at $z$=1.  Morphological complexity, such as the variation in extinction and star formation history seen in local star-forming galaxies, would be blurred out in these integrated measurements.

Gravitational lensing magnifies some distant galaxies \citep{Zwicky37,Fort88,Lynds89,Yee96}, enabling their star forming regions to be studied at higher spatial resolution than can generally be done for the unlensed population \citep{Swinbank06,Jones10,Livermore12}.  
The lensing boost also permits detection of faint optical spectral lines \citep{Hainline09,Bian10,Rigby11,Christensen12} that would be otherwise difficult or impossible to obtain, and can permit measurements of far-infrared spectral energy distributions for low--luminosity galaxies \citep{Siana08,Siana09,Rex10,Finkelstein11}.

Here, we study two galaxies that have been gravitationally lensed by the galaxy cluster Abell 2390.  We chose these two galaxies because their redshift, $z$=0.912, shifts H$\alpha$ emission into a narrowband filter on the WFC3 instrument of the \emph{Hubble Space Telescope} ({\it HST}).   The deep potential of the galaxy cluster has magnified these galaxies (by factors of 10 and 4, respectively \citep{Swinbank06}) such that their mid- to far-infrared spectral energy distributions are extremely well measured by \emph{Spitzer} and \emph{Herschel}.  Thus, for these two galaxies we have captured two seldom available quantities at $z$=1 for such low-luminosity galaxies:  the integrated H$\alpha$ luminosity and the total infrared luminosity, both of which measure galactic star formation rate.  In addition, the improved spatial resolution brought by lensing magnification allows the study of how that star formation is distributed spatially, as traced by H$\alpha$ and the rest-UV continuum.  In this paper, we explore whether conclusions based on integrated quantities like star formation rate and extinction are substantiated in two cases where lensing provides a higher spatial resolution picture of these quantities. We will relate the morphology of current star formation to the structure of the existing stellar populations.  In addition, we determine star formation rates for the two galaxies in multiple ways, and investigate how dust extinction and the spatial distribution of star formation affects the inferred star formation rates.

These two galaxies were discovered serendipitously by \citet{Pello91}.  ``Arc A'' (nomenclature following \citealt{Swinbank06}) is also known as the ``straight arc'' (component C) by \citet{Pello91} and \citet{Frye98} and as ``Abell~2390c'' by \citet{Rigby08}.
``Arc D'' (nomenclature following \citealt{Pello91} and \citealt{Swinbank06}) is also called ``Abell~2390b'' by \citet{Rigby08}.  The similar redshifts of these two arcs suggests that Abell 2390 might be lensing a group of galaxies \citep{Swinbank06}.  Arc D is a bright X-ray point source, and emits roughly equal power in the X-rays and in the aromatic features \citep{Rigby08}.  As a result, those authors concluded that both an active nucleus and vigorous star formation are present in Arc D.  The active nucleus complicates the analysis of Arc D in this work, in that there may be an unknown contribution to the star formation rates derived from H$\alpha$ and the far-infrared.  However, since both Arc A and Arc D were captured by HST in the same field of view, Arc D comes ``for free'', and it is therefore worth learning what we can from Arc D, with the significant caveat that the star formation diagnostics may be contaminated by the AGN.

Cluster lens modeling (\citealt{Pello99}; refined by \citealt{Swinbank06}) permit images of these galaxies to be reconstructed in the source plane. We incorporate previous kinematic measurements \citep{Swinbank06}, mid-infrared and submillimeter photometry, and archival {\it HST} broadband imaging. 

The structure of this paper is as follows.  In \S \ref{methods}, we discuss the observations and data reduction process, including the deprojection of images onto the source place.  In \S \ref{morphologies}, we describe the source-plane morphologies of the two galaxies in narrowband H$\alpha$ and three broadband {\it HST} filters, and relate these to their kinematics from \citet{Swinbank06}.  In \S \ref{FIR} we measure integrated star formation rates from the far-infrared in three different ways; in \S \ref{clumpsfr} we measure extincted H$\alpha$ and rest-2800 \AA\ star formation rates; and in \S \ref{sec:compare_extinction} we combine these results to constrain the ratio of stellar to nebular extinction and the total extinction in H$\alpha$ for each galaxy.  We summarize the results in \S \ref{sec:conclusions}. 

We adopt a cosmology with $H_0=71$ km~s$^{-1}$ Mpc$^{-1}$, $\Omega_m=0.27$ and $\Omega_{\Lambda}=0.73$.

When an initial mass function (IMF) is required, we adopt that of \citet{Kroupa03}; our conclusions do not depend on the choice of IMF.

\section{Methods}
\label{methods}

\subsection{{\it HST} Observations}
\label{observations}

We obtained observations with the Wide Field Camera 3 (WFC3) on {\it HST}, in two orbits of Cycle 17 guest observer time (program 11678, PI Rigby).  We integrated for 1.4~ks in the $J_{1250}$ broad-band filter, and 4.2~ks in the [Fe II] narrow-band filter.  These observations are in the infrared channel, which employs a HgCdTe detector with a $136\arcsec$ x $123\arcsec$ field of view and a plate scale of $0.1283\arcsec$/pixel.  The $19\arcsec$ separation between the two arcs in A2390 allows them to fit into a single field-of-view.  The observations were taken on 2010 September 02, with four exposures for the $J_{1250}$-band images and three for [Fe II], as summarized in Table \ref{obs}.  The first two narrow-band images were acquired during the first of two orbits, and the third narrow-band image was acquired in the second orbit along with the $J_{1250}$-band images.  Three dither positions with integer pixel offsets of $7''$ were used for each filter.

Additionally, we use {\it HST} Wide Field and Planetary Camera 2 (WFPC2) images of A2390 from the MAST public archive, from {\it HST} GO program 5352 (PI Fort) taken 1994 December 10 in Cycle 4.  The observations consist of five dither positions in the $I_{814}$-band and four dither positions in $V_{555}$-band, where the pointing changes by $1''$ between exposures.  Both arcs lie in one of WFPC2's three $150''$ x $150''$ CCDs, WF3, which has a pixel size of $0.0996''$.

\subsection{Image processing}
\label{multidrizzle}

We combined the dithered {\it HST} images for each filter using the software Multidrizzle \citep{Fruchter09}, included as part of the Space Telescope Science Institute (STScI) Python package.  This software solves several data reduction issues at once; namely, it removes cosmic rays and bad pixels while recovering a point-spread function (PSF) that is as close to Nyquist sampled as possible.  Multidrizzle shifts and rotates each dithered image to a common world coordinate system (wcs), letting the pixels `rain down' or `drizzle' onto a final grid that is more finely spaced than the original grid, where they are given a weight proportional to the area of each new pixel they cover.  Once each of the images is mapped onto the new grid, they are averaged together, rejecting cosmic rays 
and known bad pixels, to create a final image.  Optionally, the original pixels can be shrunk and given a shape, or kernel type, before they are drizzled onto the final image.  

Since this was a small program, the number of dithers was small, and did not fully sample the PSF; whole-pixel offsets were used, rather than sub-pixel offsets.  We therefore kept the parameter pixfrac set to unity.  We found little difference between the square and Gaussian kernel types, and ultimately settled on a Gaussian kernel as there were slightly fewer defects in the final images in this case.  Because we rotate our images during the drizzling process in order to match the orientation of the lensing model, it is still advantageous to use a finely spaced final grid to avoid adding shape noise to the PSF \citep{Rhodes07}.  We use a final plate scale of 0.05 arcsec/pixel.

Since the field was too crowded for background subtraction, we instead determined a median sky value for the $J_{1250}$ and F126N images in a $90 \times 40$ pixel region near the two arcs and removed this background prior to drizzling.  Because we are subtracting a flat sky, the timing of this relative to drizzling is not of particular importance; thus, we deferred the background subtraction of the $I_{814}$ and $V_{555}$ until afterwards for convenience.  Finally, we used the most recent correction tables\footnote{\href{http://www.stsci.edu/hst/observatory/cdbs/SIfileInfo/WFC3/WFC3distcoeff}{http://www.stsci.edu/hst/observatory/cdbs/SIfileInfo/WFC3/WFC3distcoeff}} to remove the geometric distortion.

\subsection{Astrometry}
\label{astrometry}

Because we aim to compare observations taken with multiple instruments over multiple epochs, it was necessary to make astrometric adjustments.  Typical absolute residual pointing errors of {\it HST} are quoted as $\sim0.2-0.5\arcsec$ in the Multidrizzle Handbook \citep{Fruchter09}, and as high as $\sim0.9\arcsec$ in the WFPC2 Instrument Handbook \citep{Heyer04}.  
We first looked for both rotation and shift between the WFC3 and the WFPC2 images, rotating the $J_{1250}$-band image by varying amounts and then cross-correlating it against the drizzled $I_{814}$-band image.  Our smallest step size of 0.001 degrees would correspond to a maximum residual shift of 0.04 pixels (0.002\arcsec) at the corner of the image, and $\sim$0.006 pixels at the position of Arc A.  After re-drizzling the $J_{1250}$-band image, we refined the translational shifts by both centroiding and cross-correlating individual stars in the images.  We find offsets of $\Delta x=5.40\pm0.07$ pixels and $\Delta y=9.32\pm0.05$ pixels between the $J_{1250}$-band and $I_{814}$-band images, which, with a plate scale of 0.05\arcsec/pixel, corresponds to a total shift of $0.537\arcsec \pm 0.015\arcsec$.  The uncertainty incorporates the centroiding error for each star, given by the rms width of the star ($\sim \textrm{its full width at half maximum}/(2\sqrt{2\ln{2}})$) divided by the square root of the number of electron counts.  We also find a rotation offset of $0.214\pm0.001$ degrees about the current image centers, which we applied to the $J_{1250}$-band and F126N images.

We next searched for a translational offset between the two narrow-band images, acquired in the first orbit, and the drizzled $J_{1250}$-band image, acquired in the second orbit.  We expect this shift to be smaller than above, as all images have the same guide stars and the typical uncertainty for re-acquisition is only 5-20 mas \citep{Fruchter09}.  We find about twice this typical value, with offsets of $\Delta x=0.69\pm0.07$ pixels and $\Delta y=0.25\pm0.05$ pixels between the drizzled $J_{1250}$-band image and the first two F126N images, for a total offset of $37 \pm 4$ mas.  In both cases, the shifts obtained from cross-correlation and centroiding of stars are similar to within 0.04 pixels. 

Combining errors leaves a total uncertainty of $\sim0.1$ pixels in the position of the F126N relative to the $V_{555}$ and $I_{814}$ images.  We translate these image-plane offset uncertainties onto the source-plane, where we find 1$\sigma$ errors of $(0.5,0.3)$ pixels near the center of Arc A with a variations of $<0.1$ pixels at other locations on the image, and $(0.6,0.5)$ pixels across the Arc D image, where the pixel scale on the source-plane is 0.005\arcsec in both cases.

\subsection{Continuum subtraction of H$\alpha$}
\label{cont sub}

In order to produce H$\alpha$ images of two arcs, we used the $J_{1250}$-band image to remove the continuum emission from the F126N image.  We scaled the $J_{1250}$-band image by the ratio of the F126N integrated filter throughput to the F125W throughput and subtracted it from the narrow-band image.  We measured the median residual flux in the continuum-subtracted image as a fraction of the flux in the $J_{1250}$-band for 67 galaxies in the field using an aperture with a diameter of 0.4\arcsec, and repeated this exercise for 88 stars with signal-to-noise ratios greater than 10, which yielded the same fractional residual to within 6\%.  The average of these two results adjusts our initial continuum scaling by 4.5\%, leaving median residual fluxes of $-0.1\%\pm 0.7\%$ for the galaxies and $0.1\%\pm 1.7\%$ for the stars.  We also note that the $J_{1250}$-band includes the H$\alpha$ emission, which introduces some uncertainty to our results; however, the continuum image is at most 10\% H$\alpha$ emission for Arc D and 7\% H$\alpha$ emission for Arc A.  We do not consider this to be a signficant effect.

\subsection{Correction for [N~II] contamination of H$\alpha$}
\label{stellarmass}

Not only H$\alpha$ but also the [N~II]~6548.1~\AA\ and [N~II]~6583.5~\AA\ lines fall into the F126N filter.  We must correct for [N~II] emission to determine the H$\alpha$ flux.  
The [N II] to H$\alpha$ ratio is used as a discriminant between star-forming galaxies and AGN \citep{BPT}.  For star-forming galaxies, the ratio is also a metallicity indicator (e.g. \citealt{PettiniPagel}), and given the existence of a mass-metallicity relation \citep{Tremonti04}, correlates with stellar mass.  We lack a spectroscopic measurement of the [N II] to H$\alpha$ ratio in these two galaxies, and instead must adopt likely values, given their properties, to correct for [N II] containation to H$\alpha$.  We thus scale from the stellar masses of these galaxies, using the z$=$1 mass--metallicity relation, to estimate the likely contribution of [N II] to the narrow-band flux.  

The stellar mass of Arc D is estimated from spectral energy distribution modeling of the observed rest-frame optical, near-infrared, and mid-infrared photometry from \citet{Swinbank06} and \citet{Rigby08} (Figure \ref{optirSED}). We use \textit{Hyperzmass} at fixed spectroscopic redshift \citep{Bolzonella00} with the updated Bruzual and Charlot SED models (CB07, \citealt{Bruzual03}), a Kroupa initial mass function \citep{Kroupa03} and Calzetti dust extinction law \citep{Calzetti00}. The SED templates have solar metallicity and the star formation histories are exponentially declining or constant. The age of the stellar population is required to be $>50$~Myr to avoid unphysically young SED models dominated by the UV emission of bright, young stars \citep{WuytsE12}. We find log(M$_*$)$=$10.9~\Msun,with a systematic uncertainty of $\sim0.4$~dex \citep{Shapley05, Wuyts07}.

For Arc A, \citet{Rigby08} were unable to measure IRAC/Spitzer photometry due to severe crowding.  The reddest band of photometry available for Arc A is observed K-band.  We therefore assume the same shape of the stellar SED for Arc A as for Arc D, and scale down the stellar mass by the ratio of the two galaxies' observed K-band fluxes  (rest 1.15~$\mu$m).  This yields a stellar mass for Arc A of log(M*)=9.8$\pm$0.4~\Msun, which is 8\% the mass of Arc D.  This quoted uncertainty does not reflect the systematic uncertainty resulting from possible differences in the stellar SED between Arc A and Arc D.

From the stellar masses, we estimate the metallicity of each galaxy using the $z\sim$0.8 mass-metallicity relation from the DEEP2 survey (\citet{Zahid11}, Equation 8). Ultimately, we use the metallicity to estimate the relative contributions of [N~II] and H$\alpha$ in the F126N filter, which we can do because [N~II]$\lambda$6583.5/H$\alpha$ is a common metallicity indicator (PP04 N2, \citealt{Pettini04}). But because \citet{Zahid11} use R$_{23}$ (KK04, \citealt{Kobulnicky04}) as their metallicity indicator, and different indicators are known to yield vastly different metallicities, we cannot directly convert the metallicity from \citet{Zahid11}'s mass-metallicity relation to an [N~II]$\lambda$6583.5/H$\alpha$ ratio. We first use the calibrations of \citet{Kewley08} (Equation 1, Table 3) to convert the KK04 metallicity to the equivalent PP04 N2 metallicity; then we use \citet{Pettini04} Equation 1 to convert the metallicity, 12+log(O/H), back to N2 $\equiv$ log(N~[II]$\lambda$6583.5/H$\alpha$).  We combine these results with the intrisinic ratio of [N~II]$\lambda$6548.1 to [N~II]$\lambda$6583.5 (\citealt{Storey00}, Table 3) and the F126N filter throughput curve, and remove the [N~II] contribution from the measured flux.  We find metallicites of 12+log(O/H) = 8.5$\pm$0.1 for Arc A and 8.7$\pm$0.04 for Arc D, which imply 20$\pm$8\% and 36$\pm$4\% [N~II] contamination for arcs A and D, respectively.

This scaling only holds if the AGN contribution to the H$\alpha$ and [N II] flux is minimal.  In Arc A this scaling seems justified, as that galaxy shows no sign of AGN activity in Chandra, Spitzer, or HST measurements \citep{Rigby08}.  For Arc D it may not be justified, as that galaxy has a known AGN.  There is no published spectroscopy of H$\alpha$ an [N II] in Arc D, from which one could estimate their AGN contribution.  If the AGN does contribute significantly to the H$\alpha$ and [N II] flux in Arc D, then the [N II]/H$\alpha$ ratio should be elevated relative to our assumption of a star forming galaxy, which would cause [N II] to be under-subtracted, and the H$\alpha$--derived star formation rate to be over-estimated in this arc.  The far-infrared star formation rates for this galaxy are also be subject to an unknown AGN contribution.

\subsection{Lensing deprojection}
\label{deprojection}

A lensing model is required to project the two arcs from the image-plane into the source-plane, thereby reconstructing their morphologies in the absence of lensing.  We use a lenging model based on that published in \citet{Swinbank06}, but revised to include additional strong lensing constraints detected in the more recent Hubble ACS image, as well as two new spectroscopic redshifts for multiple images, as described in \citet{Newman13}. 
To calculate the magnification of each arc, we transformed only the flux within the apertures shown in Figure \ref{a2390} that we later use for our image-plane photometric measurements.  The magnification $\mu$ is defined to be the ratio of lensed to unlensed flux; thus, it is not uncommon to find different values of $\mu$ for different flux distributions, as some regions of the galaxies can be more highly magnified than others.  This differential magnification is more evident for Arc A, which is closer to the brightest cluster galaxy (BCG) and more strongly lensed than Arc D.  On average, it is magnified by a factor of $\mu=11.2$, with variations of up to $\sim10\%$.  For illustrative purposes, we also calculate the magnification based purely on the change in area, and find $\mu=9.78$ in this case.  Arc D is magnified by a factor of $\mu=4.19$, with variations of less than $1\%$.  Magnifications are given in Table \ref{mags}.  For reference, Table~\ref{mags} also lists the magnification values from \citet{Swinbank06}, which differ somewhat from the values we use in this paper.  Some of this difference is due to the updates in the lensing model; some is also due to the conservative cropping that we impose to avoid contamination from cluster galaxies.

\subsection{Morphology fitting}
\label{galfit-methods}
To estimate the maximum possible contribution of a point-like nuclear source to the H$\alpha$ emission in Arc D,  we use \galfit\ Version 3.0.4 \citep{Peng02,Peng10} to fit Sersic components and PSFs in the source plane.  \galfit \ uses a Levenberg-Marquardt algorithm to find a best-fit model through $\chi^2$ minimization, convolving the model with a point spread function (PSF). 

We first construct the image-plane PSF by centering and summing stars, then map the result onto various locations in the source-plane.  We fit both the image- and source-plane PSFs with Moffat profiles in order to quantify the spatial advantage gained by lensing and to characterize the variability of the PSF in the source-plane.  The Moffat profile is commonly used to describe the PSF of undersampled images; its surface brightness profile takes the form
\begin{equation}
\Sigma(r)=\frac{\Sigma_0}{[1+(r/r_d)^2]^n}
\end{equation}
where $r_d$ is related to the full-width at half maximum (FWHM) through
\begin{equation}
r_d=\frac{FWHM}{2\sqrt{2^{1/n}-1}}
\end{equation}

In the image-plane, we keep the axis ratio fixed at unity and allow the other parameters to vary, and include a variable flat sky to remove any residual background in the PSF image.  In the source-plane, we use the best-fit central surface brightness and concentration parameter from the image-plane for that band and allow the FWHM, position angle, and axis ratios to vary.  Table \ref{psftab} shows the best-fit FWHM in each case.  Due to the nature of the reconstruction, we find that the axis ratio is no longer unity on the source-plane and the shape of the PSF changes across the source-plane image.  The second effect is much more pronounced for Arc A, which is closer to the center of the cluster and therefore more distorted than Arc D.  For the $J_{1250}$-band, we find an rms variation of $\sim0.003\arcsec$ for the major axis and $\sim0.015\arcsec$ for the minor axis, while the position angle changes by $\sim7\%$.  For Arc D, there is negligible variation ($< 1\%$) in these parameters.  We assume similar variability in the other bands.

\subsection{\emph{Herschel} photometry}
\label{sec:herschel phot}

We retrieved data from the ESA \emph{Herschel} Space Observatory \citep{Pilbratt} for Abell 2390 from the public archive.  PACS 100~$\mu$m and 160~$\mu$m data, proposal id KPGT\_dlutz\_1, were obtained as part of the PEP key program and are summarized in Table 1 of \citet{Lutz11}.  SPIRE 250~$\mu$m, 350~$\mu$m, and 500~$\mu$m data from proposal id KPGT\_soliver\_1 were obtained as part of the HerMES key program and are summarized in Table 1 of \citet{Oliver12}.  The two data sets total 7.3h of allocated time.

Following the steps outlined in \citet{Gladders12}, we performed aperture photometry on the PACS level 2 data products using an annular aperture defined by radii of 8\arcsec, 61\arcsec, and 70\arcsec.  We apply the appropriate aperture corrections from Table 1 of PACS technical report \href{http://herschel.esac.esa.int/twiki/pub/Public/PacsCalibrationWeb/pacs_bolo_fluxcal_report_v1.pdf}{PICC-ME-TN-037}.  We determine a color correction for each galaxy using Table 1 of PACS technical report \href{http://herschel.esac.esa.int/twiki/pub/Public/PacsCalibrationWeb/cc_report_v1.pdf}{PICC-ME-TN-038}; we iteratively fit the observed-frame blackbody temperature, converging on temperatures of 20 K and 19 K for arcs A and D, respectively. 

Because of crowding in the field, pure aperture photometry was not possible for the SPIRE images.  In particular, the 18\arcsec\ separation between the galaxies is smaller than the innermost radii of the apertures specified by the \href{http://herschel.esac.esa.int/twiki/pub/Public/SpireCalibrationWeb/SPIREPhotometryCookbook_jul2011_2.pdf}{SPIRE photometry cookbook}; thus, it would be impossible to determine their relative contributions to the total flux without some revisions to this method.  Instead, we used DAOPHOT, a PSF-fitting photometry routine, to detect and remove bright sources, and, when necessary, performed aperture photometry on the resulting image following the procedure from the cookbook.  We used the empirical beam profiles described in the 
\href{http://wakefield.bnsc.rl.ac.uk/icc/product_definitions/Technical_Notes/beam_release_note_v1-0.pdf}{SPIRE Beam Model Release Note V1.1}, rotated to the correct position angle for the data.  The results are summarized in Table \ref{tab: herschel phot}.

\section{Results}
\label{results}
\subsection{Multiband morphological comparison}
\label{morphologies}

We begin by discussing the morphologies of these galaxies--in particular, the relative distributions of dust and stars--to better understand how dust might complicate the measurement of their star formation rates in \S\ref{FIR} and \S\ref{clumpsfr}.  For each galaxy, three available broadband colors are influenced by both stellar age (longer wavelengths trace older stars) and dust obscuration (longer wavelengths are less attenuated).  The H$\alpha$ emission can help to disentangle these two effects; it traces the youngest stars, and while still subject to extinction, will be significantly less attenuated than the shortest wavelength broadband filter, $V_{555}$, which traces the rest-UV.  

We also aim to better understand the morphological classifications of these systems in order to put our results in context with the larger population of galaxies at $z\sim$1.  We begin that assessment here by comparing to the kinematics from \citet{Swinbank06}, and continue in a more quantitative sense in \S\ref{galfit}.

\begin{figure}[bhtp]
\begin{center}
\includegraphics[scale=0.3]{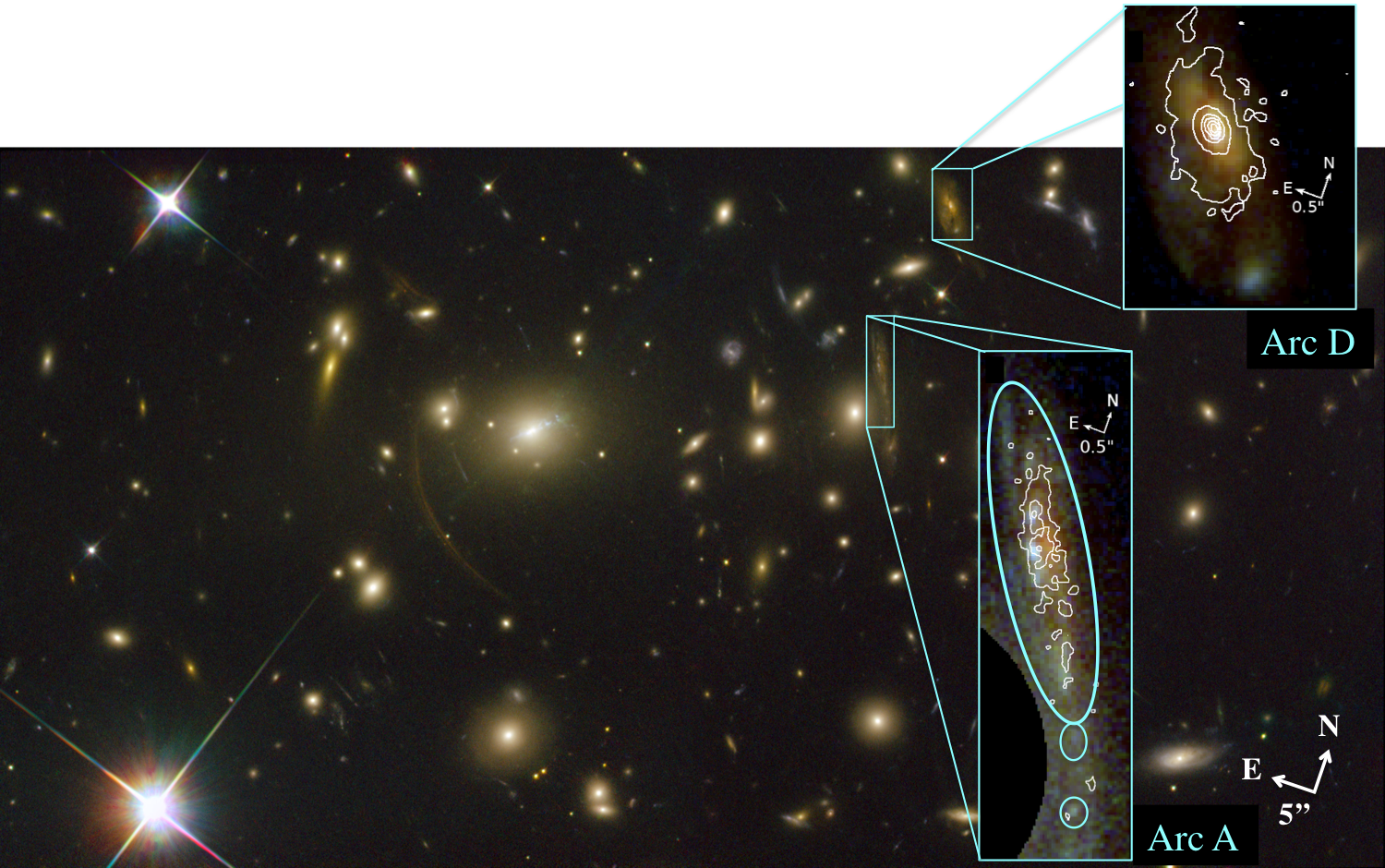}
\caption{\footnotesize Abell 2390, main photo credit NASA/ESA and Johan Richard (Caltech) ($V_{555}I_{814}I_{850}$).  Insets show $V_{555}I_{814}J_{1250}$ images of both arcs on a logarithmic scale, with H$\alpha$ contours that scale linearly from 4-13 $\sigma$ (Arc A) and 4-244 $\sigma$ (Arc D) above the background.  The ellipses on Arc A show the aperture considered for photometry.  $J_{1250}$ and H$\alpha$ images are from {\it HST}/WFC3; $V_{555}$, $I_{814}$, and $I_{850}$ are from {\it HST}/WFPC2.}
\label{a2390}
\end{center}
\end{figure}

\begin{figure}[htbp]
\begin{center}
\includegraphics[scale=0.8]{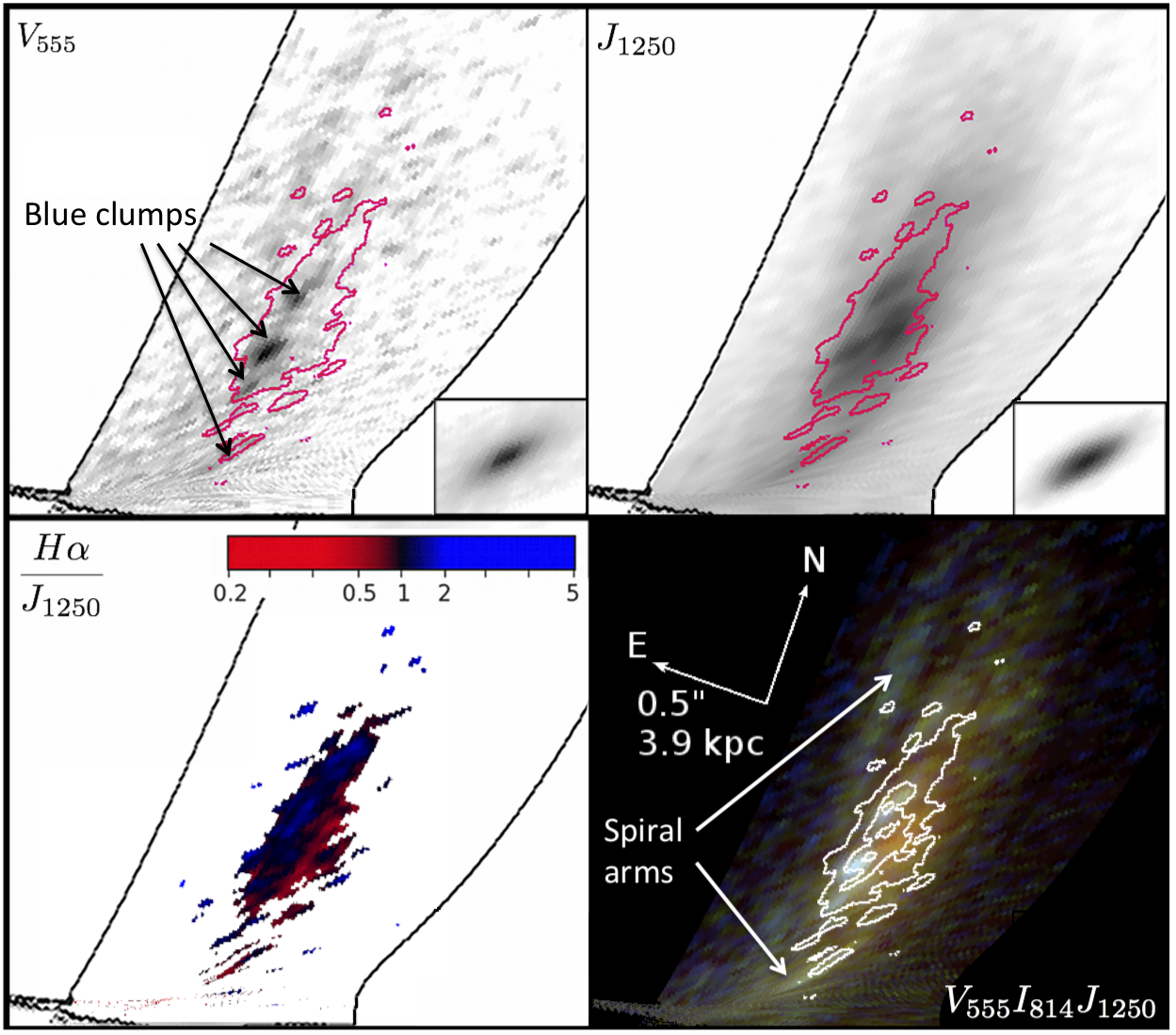}
\caption{\footnotesize Source-plane reconstructed images of Arc A.  Image scaling is same as in the inset of Figure \ref{a2390}.  H$\alpha$ contours in the lower right panel are identical to those in Figure \ref{a2390}; the top panels show only the outermost contour for reference.  The bottom right corner of each top panel shows the source-plane PSF near the center of the arc, also on a logarithmic scale.  The lower left panel shows the ratio of H$\alpha$ to the $J_{1250}$-band continuum; blue represents an excess of H$\alpha$, red represents an excess of continuum emission, and black represents the median ratio normalized to unity.  A pixel is only included in this panel if its H$\alpha$/$J_{1250}$ ratio has signal-to-noise greater than unity in the image-plane.  Solid black lines show the border of the source-plane image.}
\label{arcAsource}
\end{center}
\end{figure}

Figures \ref{a2390} and \ref{arcAsource} show the image- and source-plane images of Arc A in the $V_{555}$-, $I_{814}$-, and $J_{1250}$-bands, overlaid with contours of the H$\alpha$ flux.  There is a red clump to the west of several blue clumps, all of which are bright in H$\alpha$.  We see faint extended structure that may be spiral arms, an interpretation that would be consistent with the 2D kinematics.  Spatially resolved GMOS IFU observations of the [O II]$\lambda$3727 doublet indicate that Arc A is most likely an inclined spiral galaxy rotating with a terminal velocity $187 \pm 17\ \mathrm{km\ s^{-1}}$\citep{Swinbank06} at $\sim 9$ kpc, implying a dynamical mass of $\sim7\times 10^{10}$ \Msun.  Because the [OII]$\lambda$3727 morphology is visually similar to the observed morphologies in the images shown here, we infer that these kinematics apply to them as well.  While much of the emission drops out going from the $V_{555}$-band continuum (rest-frame 0.28~$\mu$m) to the $J_{1250}$-band continuum (rest-frame 0.65~$\mu$m), the H$\alpha$ emission appears to trace the J-band continuum fairly closely.  In contrast to the other cases, this comparison should be unaffected by extinction because the narrow-band and J-band filters are centered at the same wavelength to within a percent.

To directly compare the H$\alpha$ to the $J_{1250}$-band emission in Arc A, we show the ratio of the two source-plane images in the bottom left panel of Figures \ref{arcAsource}, where we have imposed a signal-to-noise cut of 1 for each pixel in the image-plane.  We note that a signal-to-noise cut per pixel would not be well-defined in the source-plane, as subsampling from image- to source-plane will cause this value to drop by a factor of $1/\sqrt{N_{pix}}$ for each pixel and the number of pixels subsampled by is not uniform across the image.  We see in this image that the H$\alpha$ emission is more concentrated than the $J_{1250}$-band emission at the location of the bright blue clumps, while the reverse is true for the central red region.  Quantitatively, we measure half-light radii for both the H$\alpha$ and $J_{1250}$-band images by constructing a curve of growth with elliptical apertures in the source-plane.  To correct for convolution effects, we subtract the half-light radius of the PSF from the measured half-light radii in quadrature.  We find half-light radii of $3.2\pm0.6$ kpc in H$\alpha$ and $3.9\pm0.9$ kpc at 1.25~$\mu$m measured along the major axis.

Because the image of Arc A is initially cropped in the image-plane to avoid contamination from nearby cluster member galaxies, we systematically exclude the outer regions when making these measurements.  The uncertainties in the half-light radii are derived from comparing the largest measured values, assumed to be the most accurate, to the half-light radii found using a more conservative aperture.

\begin{figure}[htbp]
\begin{center}
\includegraphics[scale=0.8]{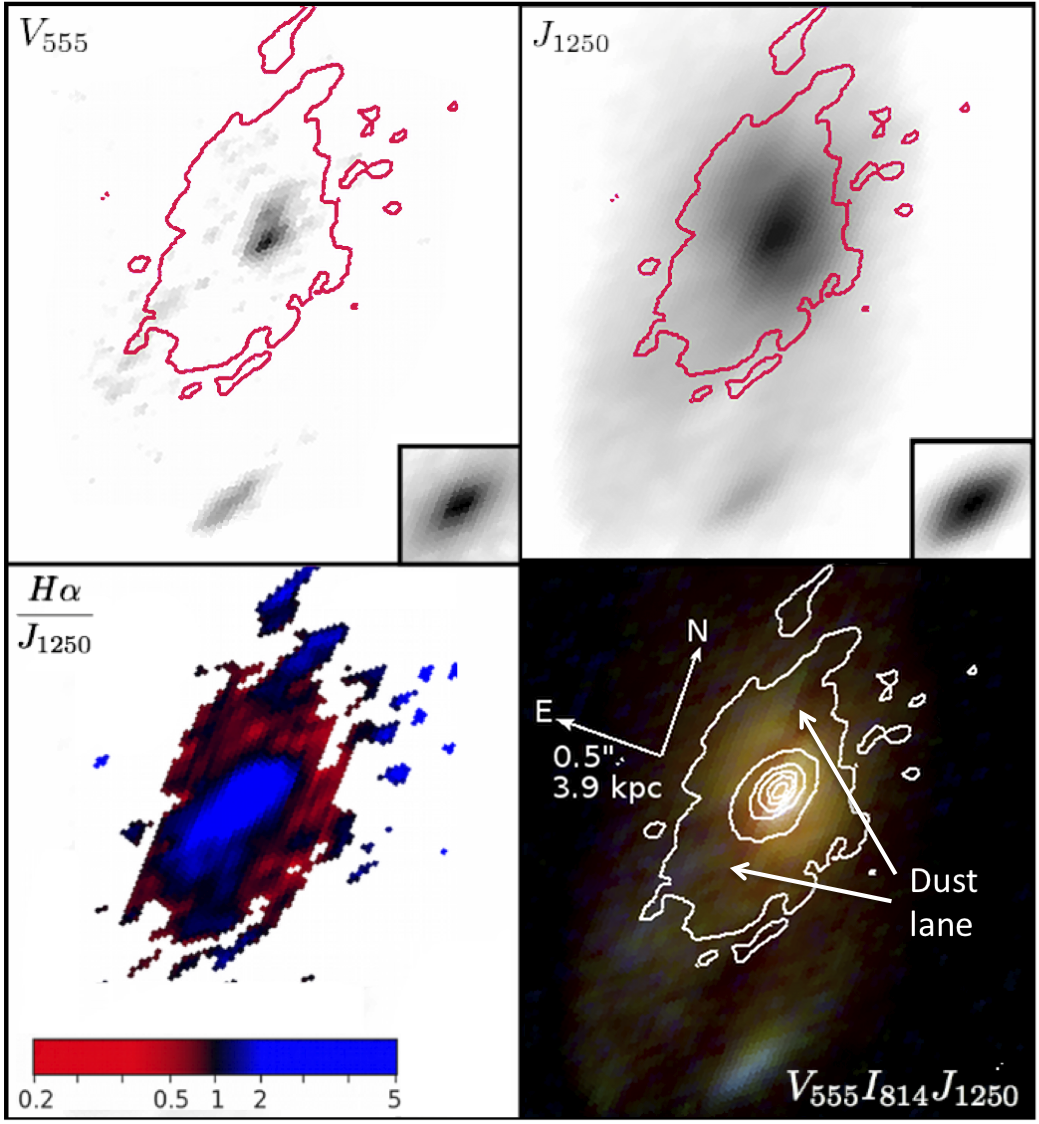}
\caption{\footnotesize Source-plane images for Arc D.  See caption for Figure \ref{arcAsource}.  Unlike Arc A, the border of the source-plane image extends outside what is shown here.}
\label{arcDsource}
\end{center}
\end{figure}

\begin{figure}[bhtp]
\begin{center}
\includegraphics[scale=0.5]{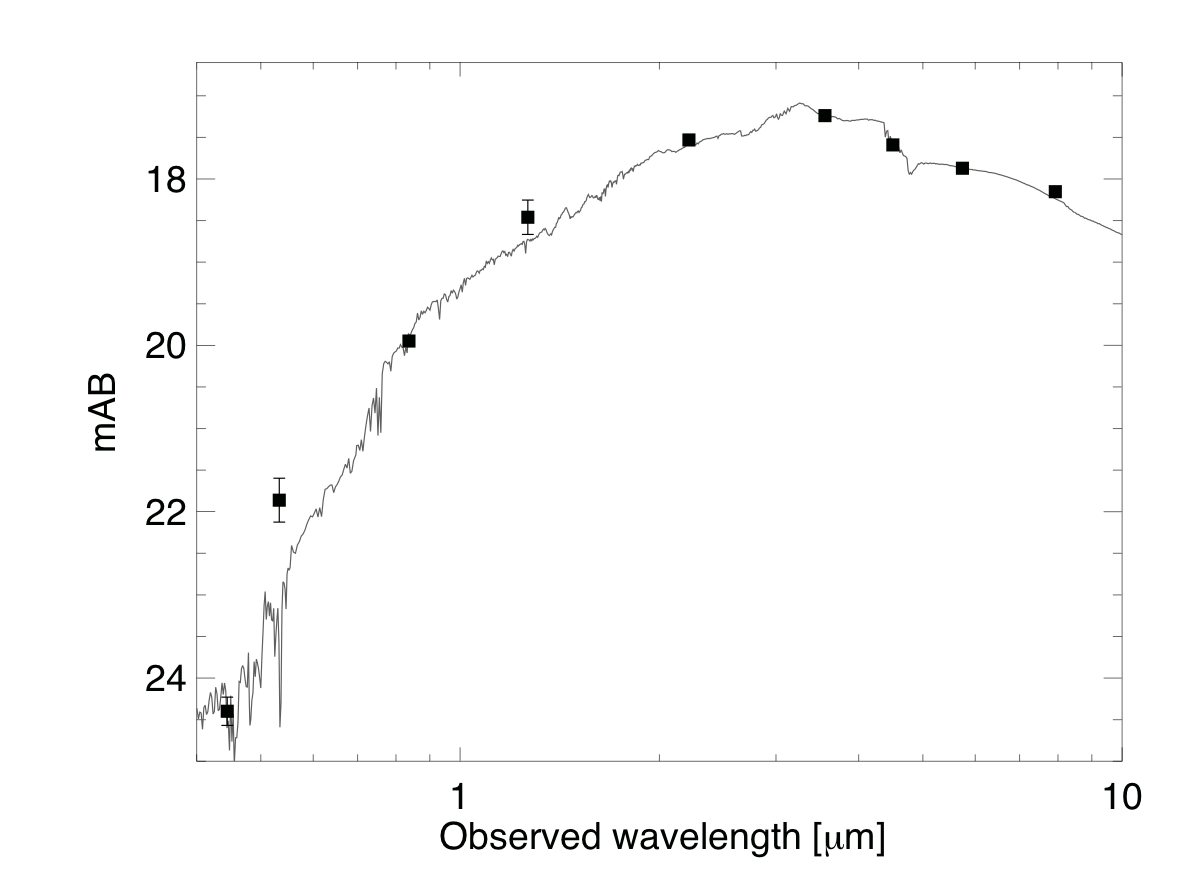}
\caption{\footnotesize Rest-optical and near-IR SED for Arc D and the best-fit Bruzual and Charlot SED model, which gives log(M*)=10.9$\pm$0.11.  All magnitudes and wavelengths are shown as observed.  Photometry is taken from \citet{Swinbank06} Table 3 and \citet{Rigby08} Table 2.}
\label{optirSED}
\end{center}
\end{figure}

In Arc D, shown in the image-plane in Figure \ref{a2390} and in the source-plane in Figure \ref{arcDsource}, most of the emission is concentrated in a $\sim5$ kpc diameter region that lies within a region of fainter extended emission.
\citet{Swinbank06} find rotation of the compact region about the north-south axis, along which we see a prominent dust lane.  The region is disturbed, with a large velocity dispersion of up to $\sim$300 km s$^{-1}$ that has two peaks, one on each side of the rotation axis.  We note that the peaks of the $V_{555}$ and $I_{814}$-band images are displaced from the peak in the H$\alpha$ image by $\sim$100 and 150 mas, respectively, (25-35 times the 1-$\sigma$ astrometric uncertainty from \S \ref{astrometry}) moving towards the western peak in the velocity dispersion.  The H$\alpha$ and $J_{1250}$-band images peak in the same location.  Some of this offset may be due to differential extinction -- the nucleus is heavily obscured by a dust lane.  The active galactic nucleus may also contribute to this offset.

The blue knot located $\sim8$ kpc to the south of Arc D in projection is suggestive of a possible recent interaction.  Although this potential companion is not visible in H$\alpha$, \citet{Swinbank06} detect it in [O II], redshifted by an additional $300\pm80\ \mathrm{km\ s^{-1}}$.  The $J_{1250}$-band flux of this knot is $\sim2\%$ the flux from the main galaxy; unlensed, it would be roughly 24th magnitude in the AB system.  While this companion is likely not massive enough to disrupt the entire galaxy, the large [OII]$\lambda$3727 velocity dispersion indicates that Arc D is not virialized.  AGN or starburst-driven winds could produce this turbulence, which would cause additional heating and ionization from shocks.  Given the large-scale complex morphology of Arc D, the velocity dispersion may instead be due to a past interaction.

We take the ratio of the H$\alpha$ to $J_{1250}$-band emission in Arc D.  The H$\alpha$ emission is more strongly concentrated than the $J_{1250}$-band continuum.  In agreement with this observation, the measured half-light radii of the H$\alpha$ and $J_{1250}$-band emission are $2.0^{+0.3}_{-0.2}$ kpc and $4.9^{+1.4}_{-0.8}$ kpc, respectively.  
It is unclear to what extent this H$\alpha$ concentration is due to the presence of an active galactic nucleus.  

We do additional Galfit modeling to constrain its maximum likely nuclear H$\alpha$ contribution, and find that a maximally bright unresolved central point source could contain up to $50\%$ of the H$\alpha$ emission.


\subsection{Integrated star formation rates from the far-infrared}
\label{FIR}
The far-infrared (8-1000~$\mu$m) luminosity, L(TIR), is widely used as an estimate of the star formation rate \citep{Kennicutt98}. Doing so requires making two assumptions:  first, that the galaxy is a good bolometer, in that most of the UV photons from young stars are absorbed by dust and re-radiated into the far-infrared; and second, that emission from dust heated by older stars is negligible.  Alternate conversions from L(8-1000~\micron) to SFR from \citet{Kennicutt98} (hereafter K98) have been proposed, which attempt to correct for dust heating by older stars, and for the fact that galaxies can be ``leaky'' rather than perfect bolometers.

The excellent spectral coverage for the two arcs studied here allows us to measure the SFR from the far-infrared in three different ways.  This allows us to gauge the importance of interstellar dust heating and non-unity covering factors in galaxies of these luminosities.

\begin{figure}[htbp]
\begin{center}
\includegraphics[scale=0.45]{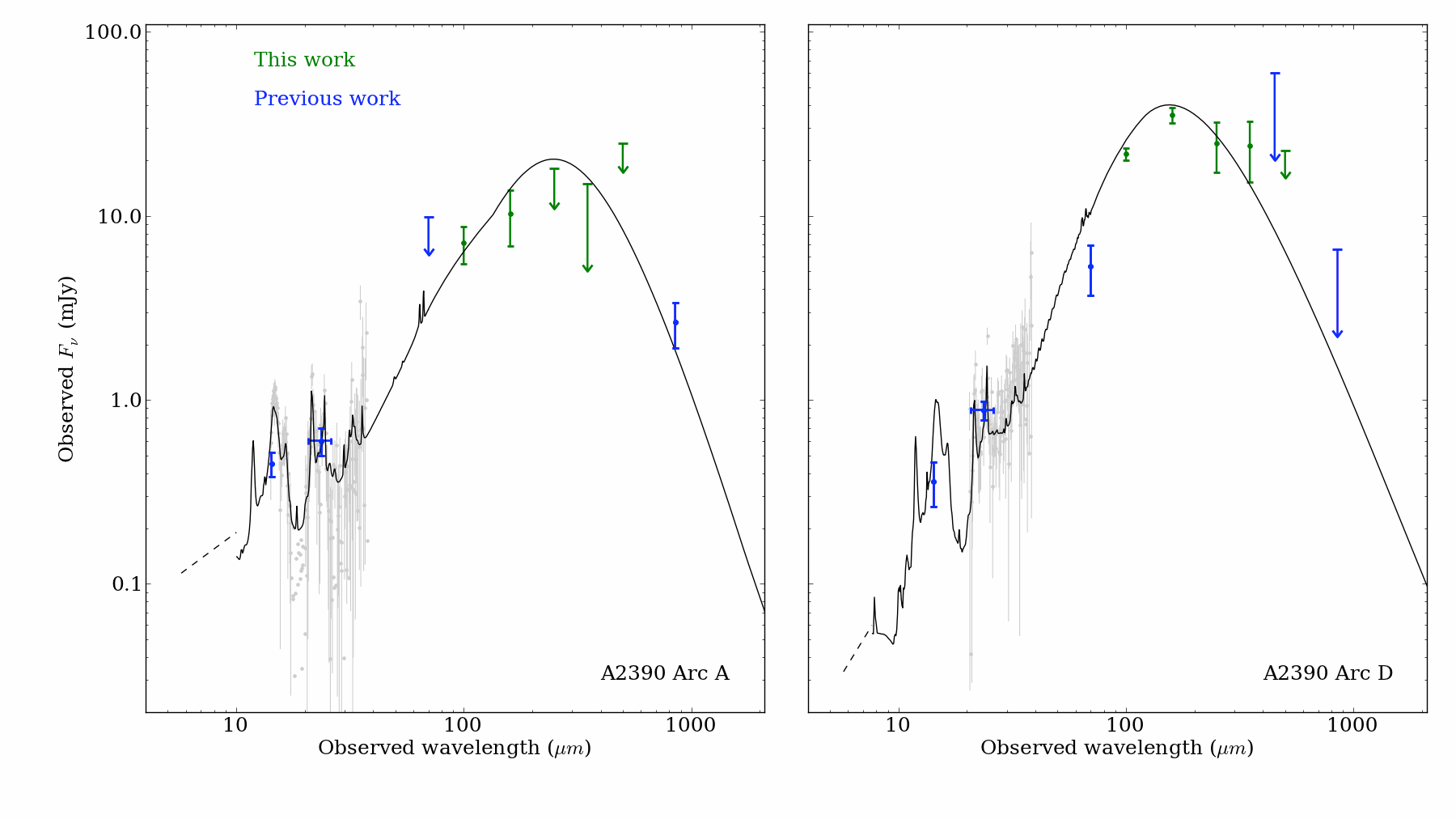}
\caption{\footnotesize Observed infrared SEDs for each galaxy.  The green points represent the \emph{Herschel} data listed in Table \ref{tab: herschel phot}, the blue points are the photometric measurements described in \citet{Rigby08}, and the black lines represent the best-fit templates from \citet{Rieke09}, which we magnify, bandwidth compress, and redshift to match the data.  We plot the Spitzer/IRS spectrum \citep{Rigby08} in gray, but exclude it from the fit due to the well-known uncertainties in flux calibrating a slit spectrograph.  3-$\sigma$ upper limits are shown with 2-$\sigma$ downward arrows.  In the case where a model lay above an upper limit, the most probable value, i.e., the flux detected in the aperture, was included in the fit, otherwise, the upper limit was ignored.  Error bars for detected points are 1-$\sigma$.  The dashed black line extrapolates the template out to rest-frame 3~$\mu$m.}
\label{SEDs}
\end{center}
\end{figure}

Previous estimates of the rest-frame total infrared ($8-1000\ \mu m$) luminosities of arcs A and D \citep{Rigby08} relied upon \emph{Spitzer}/MIPS 24 and 70~\micron\ from that paper, and submillimeter photometry from \citet{Cowie02} for Arc A and \citet{Chapman02} for Arc D.
At that time, there were no data in the $100-400\ \mu m$ observed region where the SEDs peak.  We fold in five-band $100-500\ \mu m$ \emph{Herschel} photometry from \S \ref{sec:herschel phot} and Table \ref{tab: herschel phot} to better constrain the total infrared luminosities.  Confusion noise and statistical uncertainty, quoted separately in Table \ref{tab: herschel phot}, are summed in quadrature for the purposes of the fit.  For each galaxy, we found the best-fitting \citet{Rieke09} template (Figure \ref{SEDs}), using the Levenberg-Marquardt fitting routine MPFIT \citep{Markwardt09}.  We linearly interpolate between templates of different luminosities to improve the quality of the fit: the reduced $\chi^2$ improves by factors of 1.8 (Arc A) and 1.6 (Arc D) compared to the closest provided template, and the resulting 1-$\sigma$ uncertainties in the total luminosities (Table \ref{LTIR}) are smaller than the spacing between the templates, $\Delta \mathrm{Log(}L\mathrm{(TIR)/L_\odot)} = 0.25$.

Table \ref{LTIR} shows the best-fit $L$(TIR).  Because these data are not spatially resolved, we adopt the magnification factor $\mu$ of the H$\alpha$ emission for all far-infrared wavelengths.  Since the relationship between SED shape and luminosity evolves with redshift, particularly at high luminosity \citep{Rowan-Robinson04, Rowan-Robinson05, Sajina06, Symeonidis09, Elbaz10, Hwang10}, we experimented with scaling the luminosity of the templates to find a better fit; this changed $L$(TIR) by $<$1\%.  Therefore, we did not vary the luminosity of the templates in the final fit, and instead used the fitting technique described in the previous paragraph.

For the first method of estimating the star formation rate, we apply K98 with a Kroupa initial mass function (IMF) correction, namely
\begin{equation}
\mathrm{SFR(TIR) [M_\odot\  \mathrm{yr}^{-1}]} = 0.66\times (4.5\times10^{-44})L(\mathrm{FIR)\ [erg\ s^{-1}]}
\end{equation}
to translate the far infrared luminosities into SFRs, where we have added a correction factor of 0.66 to change from a Salpeter to a Kroupa IMF \citep{Kroupa02,Rieke09}.  The Kroupa IMF has a similar slope to the traditional Salpeter IMF at the high mass end of the distribution, but drops off more steeply at the low mass end \citep{Kroupa02,Salpeter55}.  Here, L(TIR) represents the integrated luminosity from 8-1000~$\mu$m, which we obtain by numerically integrating the \citet{Rieke09} templates (Table \ref{LTIR}).  If cold dust significantly adds to the SED, we expect K98 to overpredict the SFR.

In the second method of estimating star formation rate, we address the issue of contamination from old stars using the formalism of \citet{Rieke09}, modified by \citet{Rujopakarn11b}. This formalism makes the same assumption as K98, that the far-infrared is a bolometer of all UV photons.  
What is different is that they assume that old stars may significantly contribute to these UV photons, and therefore to the far-infrared SED.  Accordingly, they use the observed-frame 24~$\mu$m flux density to trace heating from young stars only.  \citet{Rieke09} compute the ratio of L(24~$\mu$m) to L$_{tot}$(IR) for their templates as a function of $L$(TIR), and see that the ratio plateaus at $16\%$
at log ($L$(TIR)/L$_\odot$) of about 11 (see their Figure 8).  They assume that at this plateau, the 
contribution of old star heating is negligible. They then propose that L(24~$\mu$m) is a \textit{better} SFR indicator that $L$(TIR), because heating from old stars is negligible. 

 The potential pitfall of this method is that it assumes that the mid-infrared spectra of galaxies, which 
are a complex combination of continuum emission, aromatic emission, and silicate absorption, do not vary significantly from galaxy to galaxy, and are well--represented by the templates of \citet{Rieke09}.  
Theoretical work finds that the strengths of the aromatic emission features, commonly attributed to polycyclical aromatic hydrocarbons, should be sensitive to the grain size distribution and the local radiation field \citep{DraineLi2007}.

We de-magnify the observed-frame 24~$\mu$m\ flux density $f$(24~$\mu$m) from \citet{Rigby08} Table 2.  From this 24 ~$\mu$m\ flux density, we use equations 7 and 8 of  \citet{Rujopakarn11b}  
to estimate a star formation rate.

For the third method of estimating the star formation rate, we move away from the assumption that these galaxies are good bolometers, following the formalism of \citet{Kennicutt09} (hereafter K09).  They assume that H$\alpha$ photons trace unobscured emission from hot stars, and that the far-infrared traces obscured emission from hot stars.  Instead of trying to determine the star formation rate from the far-infrared alone, they add a portion of the total far-infrared to the obscured H$\alpha$ luminosity to recover the extinction-corrected H$\alpha$ luminosity.  They then turn L(H$\alpha$,corr) into a SFR using Equation 2 of K98 with a Kroupa IMF \citep{Kroupa02,Rieke09}, namely
\begin{equation}
\label{hasfr}
\mathrm{SFR(H\alpha,obs)\ [M_\odot\  yr^{-1}]} = 0.66\times (7.9\times10^{-42})L(\mathrm{H\alpha, corr)\ [erg\ s}^{-1}].
\end{equation}

In typical galaxies about half the H$\alpha$ photons go into heating dust; K09 find that this accounts for 0.24$\pm$0.06\% of $L$(TIR) (their Table 4). The fact that this calibration is empirical makes it better than the K98 formalism, which is entirely theoretical and ignores stars with ages $<$ 30 Myr.  K09 claim that in their formalism, dust heating by stars $>$ 30 Myr would contribute roughly half of $L$(TIR), i.e., interstellar dust heating is implicitly included in their Equation 1.  We integrate the \citet{Rieke09} templates\footnote{\citet{Kennicutt09} use $L$(TIR) measured over 3-1100~$\mu$m instead of 8-1000~$\mu$m.  Accordingly, to use this formalism we extrapolate the templates from 4--5~\micron\ down to 3~\micron.} and apply this method to again estimate the SFR for each galaxy.

We now compare the results of these three methods for each arc, shown in Table \ref{IRSFR}.  For Arc A, we infer a log $L$(TIR) of $10.81\pm0.08$~\Lsun\ from the 24~$\mu$m flux, compared to $10.62\pm0.05$~\Lsun\ from integrating the best-fit Rieke template from 8-1000~$\mu$m.  In other words, the 24~$\mu$m flux predicts a \textit{higher} value for $L$(TIR) than what we measure by fitting the whole FIR SED: the opposite of what we would expect if older stars contribute significantly to dust heating.  The 24~$\mu$m photometric uncertainty alone cannot explain the discrepancy.  We attribute this discrepancy to the fact that no galaxy that informed the Rieke et al.\ template set had a luminosity as low as that of Arc A. We thus conclude that Arc A is too underluminous to trust method 2.  Nor would it be prudent to use method 1 without modification, since at such low luminosities dust heating from older stars should be important.  K98 would give 
a star formation rate of $4.8\pm0.6$~\Mpyr. From Figure 8 of \citet{Rieke09}, we can estimate the fraction of the $L$(TIR) that is associated with star formation at the $L$(TIR) of Arc A: 77$^{+4}_{-1}$\%. We simply scale down the SFR from method 1 by this value, which gives $3.7^{+0.5}_{-0.4}$~\Mpyr. This agrees with the result from method 3, $3.4\pm0.7$~\Mpyr.  
Thus, for Arc A, method 1 (after a correction for interstellar dust heating) and method 3 yield the same star formation rate, and method 2 should not be used since it was constructed for more luminous galaxies.

For Arc D, the SFR inferred from the FIR (method 1) is SFR(TIR)=$29\pm1$~\Mpyr.  This is close to 
the value inferred from the value extrapolated from 24$\mu$m (method 2); SFR(24$\mu$m)=$24\pm3$~\Mpyr.  This implies that the SED of Arc D is well-described by the \citet{Rieke09} templates, and that dust heating by old stars is negligible, neither of which are surprising at these luminosities.  However, the K09 method of summing a fraction of the FIR and the obscured H$\alpha$ luminosity (method 3) yields a SFR of SFR(H$\alpha$+TIR)=$19\pm 3$ \Mpyr, which disagrees with method 1 at 3$\sigma$.  We speculate that the poor performance of the K09 method may not be surprising, since the K09 sample included only a few galaxies as luminous as Arc D.  By contrast, the \citet{Rieke09} templates are based on galaxies whose luminosities are similar to Arc D.  The X-ray--bright AGN in Arc D may also contribute to both the FIR and H$\alpha$ flux, throwing off the star formation rate methods to differing degrees.

\subsection{Star formation rates in clumps}
\label{clumpsfr}
Now that we have estimated the total SFR for each galaxy from the far-infrared luminosity, we measure the extincted H$\alpha$ and UV SFRs to constrain the relative amounts of extinction to the nebular regions and to the hot stars that generate the UV flux.
Because we aim to compare these results to other $z\sim1$ studies at lower spatial resolution \citep{Garn10}, we first measure integrated quantities to give us a basis for discussion.  Since lensing magnification lets us resolve spatially distinct clumps in Arc A, we then decompose its extincted SFR measurements into separate components, and re-examine the results.  In this way, we can test whether integrated measurements of the stellar to nebular extinction ratio are robust, or if we lose information by neglecting differential extinction.  

We use simple aperture photometry to measure the integrated H$\alpha$ and rest-UV SFRs for both galaxies, as summarized in Table \ref{hauvsfrtab}.  Equation \ref{hasfr} is used to turn H$\alpha$ luminosities into extincted SFRs.  For the 2800 \AA rest-UV, we use \citet{Kennicutt98} with a Kroupa IMF correction,
\begin{equation}
\label{uvsfr}
\mathrm{SFR(UV,obs)\ [M_\odot \  yr^{-1}]} = 0.66\times (1.4\times10^{-20})\frac{\lambda^2}{c} L_\lambda(\mathrm{UV)\ [erg\ s^{-1}\ \AA^{-1}]}
\end{equation}
where $\lambda$ is the rest wavelength in cm, and $L_\lambda$ is the luminosity density corrected for bandwidth compression, to turn luminosities into SFRs.  (For convenience, we have converted from $L_\nu$ to $L_\lambda$.)

\begin{figure}[htbp]
\begin{center}
\includegraphics[scale=0.5]{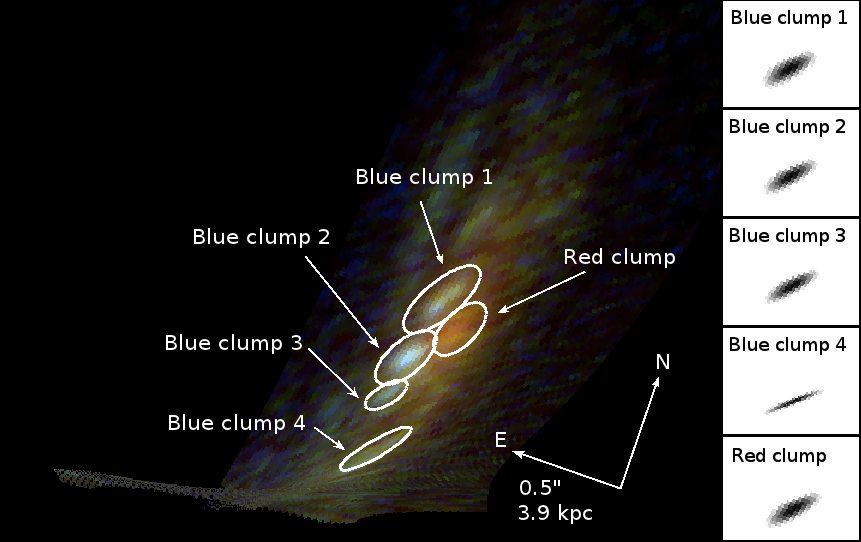}
\caption{\footnotesize Apertures used for clump photometry of Arc A, overlaid on the source-plane $V_{555}I_{814}J_{1250}$ color composite image.  On the right is the $J_{1250}$-band PSF as it appears at the location of each clump on the source-plane.}
\label{arcAclumps}
\end{center}
\end{figure}

We then take advantage of high spatial resolution and measure H$\alpha$ and rest-UV fluxes for individual bright clumps in Arc A, as indicated in Figure \ref{arcAclumps}.  Extincted source-plane SFRs are tabulated in Table~\ref{hauvsfrtab}.   Our clump definitions are based on the visual appearance of the broadband image, and differs from that in \citet{Livermore12}, where the H$\alpha$ emission is used to define clumps.  Here, clumps that are too faint or too dusty to show up prominently in the rest-UV image are included with the diffuse emission instead, with the exception of the red clump which coincides with the dynamical center of the galaxy.  To determine aperture corrections, we transform the PSF within a circular 2\arcsec\ radius aperture on the image-plane onto the source-plane in the appropriate locations, and compare the flux enclosed within the desired clump aperture to the total flux of the PSF.  
Because several of clumps are adjacent to one another, we make a geometric correction that avoids overestimating their total flux.  Typically, we find that about 3-10\% of the total emission from a clump will overflow into a nearby aperture.  
From here, we calculate SFRs from the flux of each clump.  

The results are summarized in Table \ref{clumpphot}.  In H$\alpha$, $40\pm1\%$ comes from small ($\lesssim 1$ kpc) clumps, with $31\pm 1\%$ coming from the blue clumps and $9\pm1\%$ from the red clump; in the rest-UV, $33\pm3\%$ of the observed emission is from the blue clumps and $\lesssim 6\%$ is from the red clump.  The percent of H$\alpha$ emission in clumps is intermediate between that of the $z\sim0$ SINGS sample at 31\%, and the $z~1-1.5$ lensed galaxies of \citet{Livermore12} at 50\% \citep{Kennicutt03,Livermore12}.  This is consistent with the trend that higher $z$ galaxies tend to be ``clumpier'' (e.g., \citet{Livermore12,Forster-Schreiber11,WuytsS12}) than galaxies at lower $z$.

Since Arc D does not have clumps, it cannot be analyzed in this way. 

\subsection{Comparing nebular and stellar extinctions}\label{sec:compare_extinction}
Several studies \citep{Forster-Schreiber09, Garn10, WuytsS11a, Calzetti00} have found that nebular lines are typically more extincted than the stellar continuum.  \citet{Calzetti01} suggest that the emitting gas is only found in dusty, star-forming regions, where there are $<$ 20 Myr old stars to ionize it.  Over time, stars will start to diffuse out of their birthplaces and into the rest of the galaxy, where there is less dust \citep{Calzetti01}.  Thus, in this picture, the H$\alpha$ emission should be more attenuated relative to the UV than we would expect from a reddening law alone.  

Fortunately, we can estimate the total unextincted SFR with data in the far-infrared, where dust grains thermally re-emit the energy they gain from absorbing UV photons.  We will then find and compare the attenuated rest-H$\alpha$ and -UV SFRs to constrain how much extra dust the ionized (nebular) gas could see compared to the stars.

We now compare the extincted H$\alpha$ and UV SFRs of the previous subsection to the total, unextincted SFR from the far-infrared (\S \ref{FIR}), to estimate the relative amounts of extinction of the nebular emission and the stars.  We do this for both integrated quantities, where we must assume the dust is uniformly distributed across each galaxy, and for the separate clumps in Arc A, where each component can see a different amount of dust.

We first find the attenuation in H$\alpha$ as a function of the difference in extinction between the stellar component at 2800 \AA and the nebular component at the wavelength of H$\alpha$, and the ratio of the observed star formation rates.  Because the unextincted SFR should be the same no matter what indicator we are using, we can write
\begin{equation}
\mathrm{A(H\alpha)-A(UV)}=-2.5 \log10 \left( \frac{\mathrm{SFR(H\alpha,obs)}}{\mathrm{SFR(UV,obs)}}\right)
\label{AHaminusAUV}
\end{equation}
where A(H$\alpha$) and A(UV) represent the magnitudes of extinction at H$\alpha$ and 2800 \AA, and SFR(H$\alpha$) and SFR(UV,obs) are the observed SFRs calculated above.  To allow freedom in the choice of reddening law, we express our equations in terms of an extinction ratio, A(UV)/A(H$\alpha$), which encompasses both the reddening law and the ratio of stellar to nebular extinction. 

Re-arranging (\ref{AHaminusAUV}) yields:
\begin{equation}
\mathrm{A(H\alpha)}=\dfrac{2.5}{\mathrm{A(UV)/A(H\alpha)}-1} \log10 \left( \frac{\mathrm{SFR(H\alpha,obs)}}{\mathrm{SFR(UV,obs)}}\right)
\label{AHa_HaUV}
\end{equation}
For a given A(H$\alpha$), the inferred corrected SFR, SFR(H$\alpha$,corr) takes the form
\begin{equation}
\mathrm{SFR(H\alpha,corr)=SFR(H\alpha,obs)10^{A(H\alpha)/2.5}}
\label{SFRcorr}
\end{equation}
 
\begin{figure}[htbp]
\begin{center}
\includegraphics[scale=0.4]{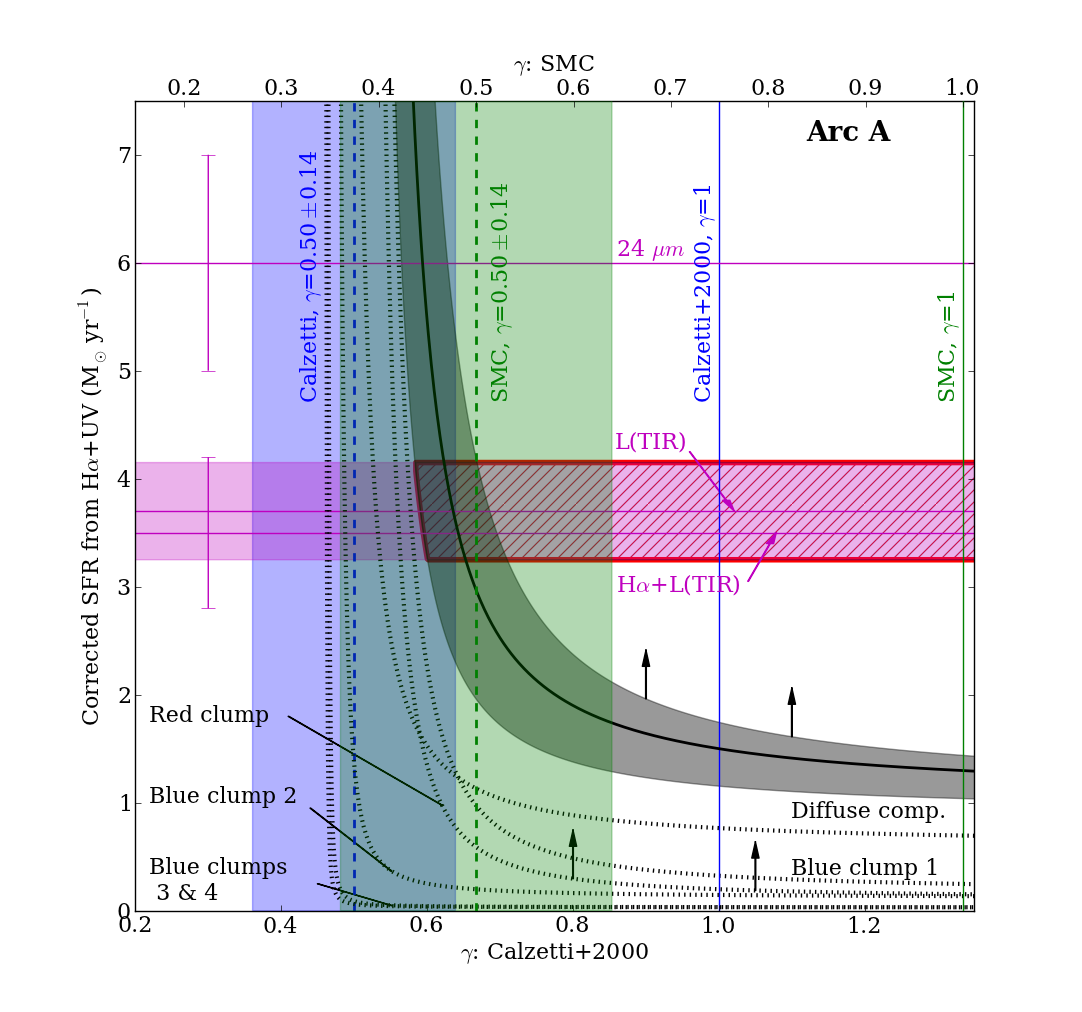}
\caption{Constraints on the ratio of the stellar extinction to the nebular extinction for Arc A.    
Each clump we have defined is treated separately, as is the diffuse component (black dotted lines).
The solid black curve shows the sum of the extinction-corrected star
formation rates (SFRs) as a function of $\gamma$, where $\gamma$ is
the ratio of stellar to nebular extinction at a given wavelength,
$A(\lambda)_{stellar}=\gamma A(\lambda)_{gas}$.  
The bottom axis
assumes a \protect\citet{Calzetti00} reddening law, while the top axis
assumes an SMC reddening law \protect\citep{Bouchet85, Prevot84}.  For
each value of $\gamma$, the observed H$\alpha$ and UV fluxes determine
the SFR represented by the curve and its 1-$\sigma$ uncertainty (in
grey).  The horizontal magenta lines show the values of the SFR
inferred from the three methods that rely on the far-infrared.
Vertical error bars show the 1-$\sigma$ uncertainty for methods 1 and
3; the magenta shaded region shows the 1-$\sigma$ uncertainty for
method 2.  For reference, vertical blue (Calzetti) and green (SMC)
dotted lines and shading show the values of $\gamma$ observed
by \protect\citet{Garn10} of $0.5 \pm 0.14$.  Solid vertical lines
represent $\gamma=1$, in which there are equal amounts of extinction
for the stellar and nebular emission.  The right-hand panel shows Arc
A, following the same conventions as the left-hand panel.  The
extinction-corrected SFR for the red clump is a lower limit, making
the summed SFR from the H$\alpha$ and UV fluxes a lower limit as well.
The region outlined in red shows the 1-$\sigma$ overlap between the
far-infrared method 2 and the H$\alpha$+UV method of determining the
SFR, and implies $\gamma$ $\gtrsim$ 0.64 (0.48) assuming a Calzetti
(SMC) reddening law. See \S~\protect\ref{sec:compare_extinction}.
}
\label{SFRvred}
\end{center}
\end{figure}

In Figure \ref{SFRvred} we plot the relation between the inferred SFR(H$\alpha$,corr) and A(UV)/A(H$\alpha$).  For Arc A, we show each clump with a different A(H$\alpha$) and sum the components to get the total, corrected SFR.  Moving to the right on the plot, the SFR (shown by the black curves) decreases as the extinction ratio increases.

Since the extinction law of $z\sim1$ galaxies has not been definitively established, we use two extinction laws that span the range of observed diversity of star-forming galaxies in the nearby Universe, namely, the extinction law of \citet{Calzetti00} which is based on nearby starburst galaxies, and the extinction law for the SMC \citep{Bouchet85, Prevot84}.  The blue vertical line shows A(UV)/A(H$\alpha$) $=$ 2.19, as predicted by the Calzetti extinction law if we assume that the stellar and ionized gas components see an equal amount of dust.  The green line shows the extinction law for the SMC, at A(UV)/A(H$\alpha$)=2.93.  

As mentioned earlier, previous results \citep{Forster-Schreiber09, Garn10, WuytsS11a, Calzetti00} find that the nebular emission is more extincted than stars in star-forming galaxies.
Following the convention of \citet{Garn10}, we let $\gamma$ be the ratio of stellar to nebular extinction, so that A($\lambda$)$_{stellar}$ = $\gamma$A($\lambda$)$_{gas}$.  \citet{Garn10} find $\gamma$ = 0.50$\pm$0.14 in their sample of $\sim500$ galaxies, slightly higher than the 0.44$\pm$0.03 used by \citet{Calzetti00} based on a sample of only 19 galaxies \citep{Calzetti97}.  The blue and green dashed lines and stripes show A(UV)/A(H$\alpha$) for $\gamma$=0.50 and its 1-$\sigma$ spread.  \citet{Garn10} only consider a \citet{Calzetti00} law; we assume a similar distribution of $\gamma$ would hold under an SMC-like law.  

We now fold in the extinction-free star formation rate measured from the far-infrared luminosity to constrain $\gamma$.  The horizontal magenta lines/stripes show results from each of the three far-infrared methods to determine SFR from  \S\ref{FIR}.
The intersection between the values allowed by the total H$\alpha$ and UV observations (the solid black curves) and the far-infrared observations (the magenta lines) yields A(UV)/A(H$\alpha$), and thus $\gamma$.  We highlight the 1-$\sigma$ allowed values in red, where use the preferred far-infrared SFR values from \S \ref{FIR}.

We now analyze Arc A.  Because Arc A has distinct clumps of UV and H$\alpha$ emission, we allow each of the components (Figure \ref{arcAclumps}) to have a different A(H$\alpha$).  For simplicity, we assume that the same extinction ratio describes every component.  We sum the individual extinction-corrected SFRs (dotted lines in Figure \ref{SFRvred}) to get a total SFR curve.  This gives $\gamma$ $\gtrsim$ 0.64 (0.48) under the Calzetti (SMC) extinction law, using the Kennicutt methods to estimate SFR from the far-infrared.  If we were to treat Arc A in an integrated sense and use a single A(H$\alpha$) for the whole galaxy, we would find $\gamma$=0.59$\pm$0.05 (0.44$\pm$0.03).  

\begin{figure}[htbp]
\begin{center}
\includegraphics[scale=0.4]{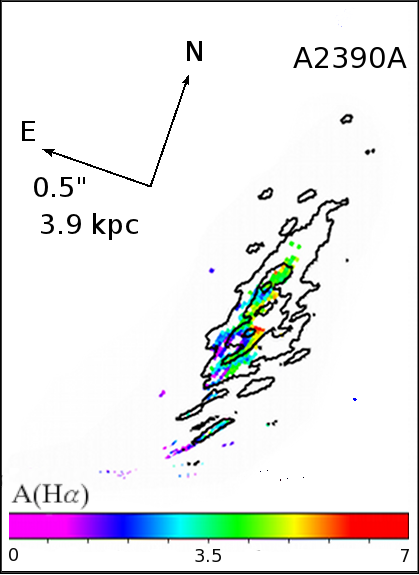}
\caption{\footnotesize Map of A(H$\alpha$) for Arc A, from the ratio of SFR(H$\alpha$,obs)/SFR(UV,obs), using a \citet{Calzetti00} reddening law and assuming the best-fit stellar to nebular extinction ratios as discussed in \S \ref{sec:compare_extinction}: $\gamma$=0.63 for Arc A.  This is a lower limit on $\gamma$, or a lower limit on A(H$\alpha$).  Only pixels with signal-to-noise $\geq0.6$ in the image-plane are included.  H$\alpha$ contours are shown for reference.}
\label{AHaimage}
\end{center}
\end{figure}

However, because no UV flux is detected from the nucleus (the red clump) of Arc A, its extinction, and therefore its amount of obscured star formation, is unconstrained.  Figure \ref{AHaimage} emphasizes this: the blue clumps of Arc A are almost dust-free, while the red clump is too dusty to show up after a signal-to-noise cut.  If we set $\gamma$ = 1 (the ionized gas and stars have the same extinction), we can calculate an upper limit for A(H$\alpha$) in the nucleus.  In this case, there would need to be 2.4 $\pm$ 0.5 (2.6 $\pm$ 0.5) M$_\odot$ yr$^{-1}$ of star formation hidden within the red clump to reach the total SFR predicted by L(TIR).  When we compare to the observed SFR in the red clump from Table \ref{clumpphot}, 0.09$\pm$0.01 M$_\odot$ yr$^{-1}$, equation (\ref{SFRcorr}) tells us it would need to have A(H$\alpha$) = 3.4 $\pm$ 0.3 (3.6 $\pm$ 0.2) to account for the missing star formation.  Going back to our lower limit, equations (\ref{AHa_HaUV}) and (\ref{SFRcorr}) with $\gamma$ = 0.64 (0.48) tell us that the red clump must have A(H$\alpha$) $\gtrsim$ 2.1 already.  Therefore, if the true extinction A(H$\alpha$) in the nucleus were to be $\sim$ 1.3 (1.5) magnitudes higher than our lower limit, this would nicely account for the far-infrared--inferred star formation rate, without invoking a different amount of extinction for the ionized gas compared to the stars.  Thus, the hypothesis that the stars and the ionized gas are well-mixed and experience identical extinction values is consistent with the spatially--resolved data.  That hypothesis would have been improperly ruled out from the integrated measurements, revealing the limitations of calculating an ``average extinction'' for a heterogeneous galaxy. 

The presence of an AGN in Arc D invalidates the assumptions of this analysis, so we do not perform it.

\begin{figure}[htbp]
\begin{center}
\includegraphics[scale=0.4]{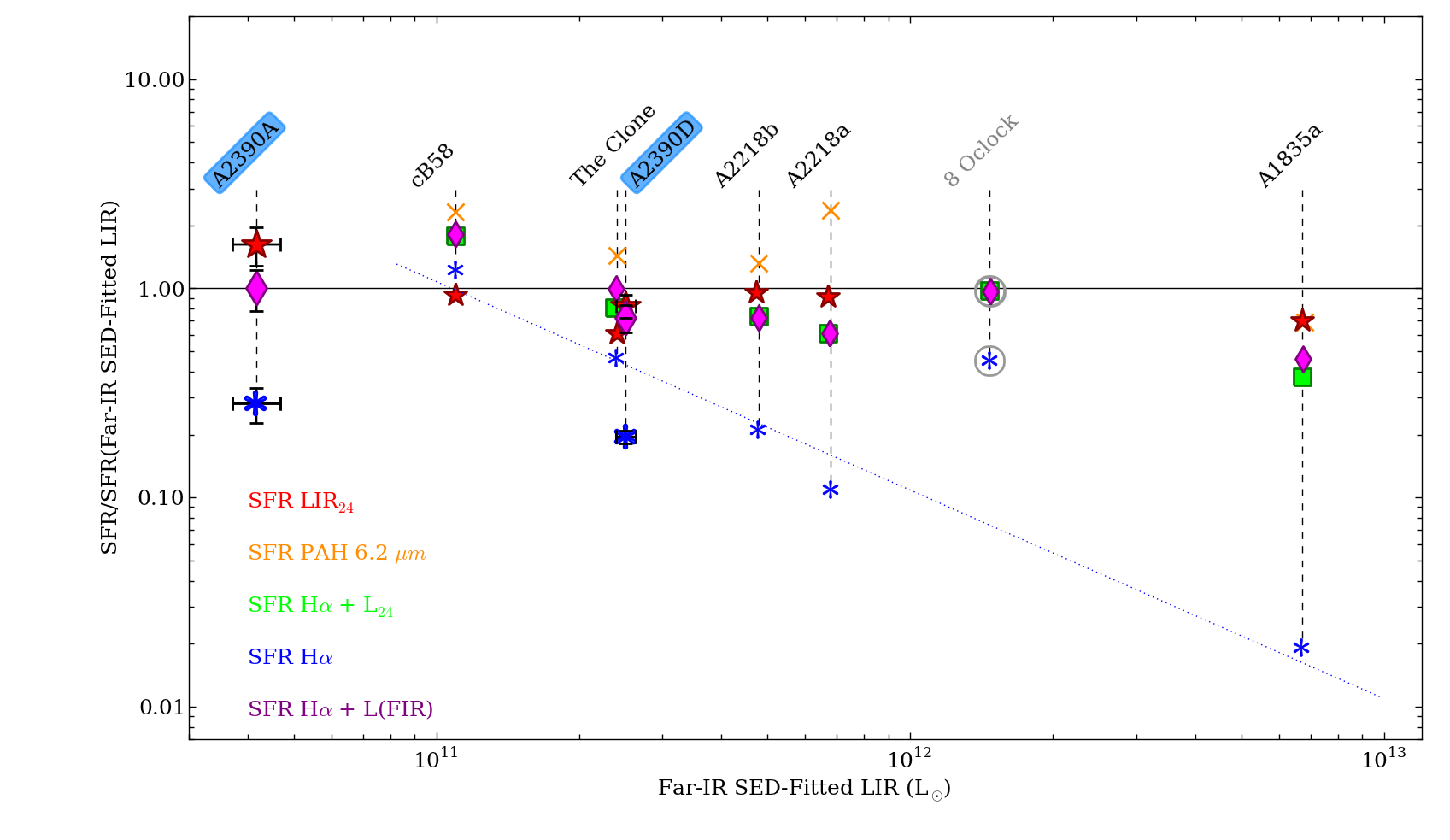}
\caption{\footnotesize Left-hand panel of Figure 8 from \citet{Rujopakarn12}, which compares the SFR measured using the specified SFR indicator to the SFR inferred from the 8-1000~$\mu$m luminosity obtained through fitting the infrared SED of each galaxy (the K98 method).  We add our results for arcs A and D, highlighted in blue.  For Arc A, we include the 13$^{+1}_{-4}$\% correction for cold dust heating discussed in \S \ref{FIR}.  The red stars show the far-infrared SFR from the observed 24~$\mu$m flux \citep{Rujopakarn11b}; while the purple diamonds show the far-infrared SFR from the K09 formalism.  The blue points represent the H$\alpha$ SFR before extinction correction; we expect these points to have ratios $< 1$.  The original objects are lensed galaxies at $1<z<3$, and are described in \citet{Rujopakarn12}.}
\label{rpkn}
\end{center}
\end{figure}

We briefly contextualize these H$\alpha$ extinction measurements in terms of the galaxy luminosities.  In our Figure \ref{rpkn} we reprint Figure 10 of \citet{Rujopakarn12}, adding Arc A and Arc D.  Though Arc D has an unknown contribution from an AGN to its far-infrared and H$\alpha$ luminosity, it does not stand out in this diagram, and appears to be a typically dusty LIRG.
However, Arc A has a much higher H$\alpha$ extinction than the trend of \citet{Rujopakarn12} would predict for a galaxy of such low luminosity.  By contrast, \citet{Garn10} predict higher extinction at L(IR)$<10^{11}$~\Lsun, values consistent with our observables. Thus, Arc A serves as an example that low-luminosity galaxies can contain significant dust and obscured star formation.

The \citet{Calzetti00} was built from integrated measurements of starburst galaxies.   As such, it is bluer than other reddening laws because some of the light escapes relatively easily, while light from deeply embedded regions is more reddened \citep{Calzetti01}.  This is similar to what we observe here: the UV emission is stronger than we'd expect based on only the \citet{Calzetti00} reddening curve.  Two obvious explanations for this result present themselves.  One, the ionized gas is more obscured than the stars:  $\gamma$ $<$ 1.  Two, parts of the galaxy are heavily obscured.   Extra dust could lie in the reddened nucleus of the galaxy, as we just discussed, or in deeply embedded regions of the blue, star-forming clumps.  The current data cannot differentiate between these possibilities.  The constraints are worse in Arc A because the dustiest regions are not detected in the UV; by contrast, in Arc D, even the dustiest regions are UV detections.

A still more extreme scenario is that the H$\alpha$ and UV emission could be spatially unrelated to the far-infrared.  This could arise if there is little dust in the H$\alpha$-bright regions, while the far-infrared regions are too heavily obscured to allow detectable H$\alpha$ to escape.  Throughout this analysis, we have assumed that the TIR, H$\alpha$, and UV all measure the same quantity, the star formation rate, though they are affected to varying degrees by extinction.
It is possible that these diagnostics measure spatially distinct regions with radically different extinction.  Unfortunately, this hypothesis cannot be tested even with the high spatial resolution and narrow-band H$\alpha$ imaging of this program.  Instead, high spatial resolution imaging in the extinction--robust far-infrared would be required.

\section{Summary \& Conclusions}\label{sec:conclusions}
In this paper, we have used the magnification boost of gravitational lensing to study in detail the star formation in two galaxies at $z$=0.912.  We have used two sets of diagnostics of star formation that are not generally available for galaxies of these luminosities at these redshifts:   narrow-band H$\alpha$ images and high-quality 24--1000~\micron\ spectral energy distributions.  We have constrained the total star formation rates from the far-infrared, and then compared to the measured star formation rates from the UV and from H$\alpha$ to constrain the extinction law and the relative amounts of extinction in the stars and the ionized gas.  The broad-band Hubble images show that these two galaxies have very different morphologies of star formation: one a nuclear starburst, the other comprised of several UV-bright clumps.  

Even though these two lensed galaxies behind A2390 likely live in the same group at $z$=0.912, their physical characteristics are quite different. Arc A, morphologically speaking, looks like other rotationally-supported clumpy galaxies at slightly higher redshifts, $z\sim$1-3 (e.g., \citet{Forster-Schreiber09,WuytsS12}).  Its infrared luminosity, $(4.2\pm0.5)\times 10^{10}$\Lsun, is more than a factor of 2 below L* at $z\sim1$ \citep{Caputi05}; as such, were it an unlensed galaxy, it would not have been detected by \emph{Spitzer} at the depth of the mid-infrared deep surveys (c.f. \citealt{Le-Floch05}, their Figure 9).  Thus, Arc A is a chance to study a low-luminosity galaxy at $z\sim1$.  The  \citet{Rieke09} templates do not describe the SED of Arc A well; the 24~\micron\ flux overpredicts the measured total infrared luminosity. We speculate that this may be due to the low luminosity of Arc A.
Integrated star formation rates return a ratio of stellar to nebular extinction of $\gamma =0.59\pm 0.05$ for a Calzetti reddening law and 0.44$\pm$0.03 for an SMC reddening law.  This also agrees with the distribution seen by \citet{Garn10}.  However, examination of the star formation and extinction on a clump-by-clump basis reveals the limits of such integrated quantities.  The nucleus has only a lower limit on its extinction, since it was not detected in the rest-UV.  If the nucleus contained an extra 1.4 magnitudes of extra extinction in H$\alpha$, then the inferred ratio of stellar to nebular extinction would be unity.  Therefore, the enhanced resolution and depth of the observations of this lensed galaxy reveal the limitations of comparing spatially--integrated extinction measurements.  Perhaps obscured nuclear star formation such as seen in Arc A are common features in star-forming galaxies at these redshifts. 

Arc D's disturbed morphology and kinematics suggest that it had a recent interaction.  A large dust lane runs across it.  Its infrared brightness and star formation rate of 24-28 M$_\odot$ yr$^{-1}$ qualify it as one of the ``luminous infrared galaxies'' that dominate the global SFR at $z\sim$1 \citep{Le-Floch05}.  It is well-fit by the \citet{Rieke09} templates, and its star formation rate is sufficiently high that dust heating by old stars is negligible.  Based on its inferred stellar mass we predict it has roughly solar metallicity.
The H$\alpha$ emission in Arc D is concentrated in the nucleus, and is not as extended as the emission from older stars.  This may well be due to the luminous AGN within Arc D, indicated by bright X-ray emission \citep{Rigby08}.

Our results show the potential of gravitationally lensed systems for the study of distant galaxies at $z\sim$1 and beyond.  The high magnification provided by lensing clusters enables us to probe background galaxies on small, sub-kiloparsec scales relevant for star formation.  In this study, we pushed the limits of the available data to see what we could learn about these two galaxies given the advantage of lensing combined with {\it HST} narrow-band H$\alpha$ imaging.  In the future, better statistics and resolved far-infrared observations of gravitationally lensed systems could expand on these results and ultimately help us to understand of the morphology of star formation in the high-redshift Universe.

\section{Acknowledgments}
We thank B. Koester and E. Wuyts for helpful discussions, and J. Richard for constructing the lens model.
S.V. acknowledges support from NASA through a Senior NASA Postdoctoral Program (NPP) award held at the NASA Goddard Space Flight Center, and from the Alexander von Humboldt Foundation for a ``renewed visit'' to MPE Garching in 2012.
We made use of Wright's cosmology calculator, \citet{Wright06}.
Support for program \# 11678 was provided by NASA through a grant from the Space Telescope Science Institute, which is operated by the Association of Universities for Research in Astronomy, Inc., under NASA contract NAS 5-26555.

\bibliography{mybib}

\section{Tables \& Figures}

\begin{table}[htbp] \footnotesize
\begin{tabular}{l l c c c c}
  \hline
  Filter & Instrument & UT Date & Total integration time (s) & Pivot $\lambda$ (\AA)\\
  \hline
  F125W (J) & WFC3/IR & 2010-09-02 & 1397 & 12490\\
  F126N & WFC3/IR & 2010-09-02 & 4198 & 12590\\
  F814W (I) & WFPC2/WF3 & 1994-12-10 & 10500 & 8336\\
  F555W (V) & WFPC2/WF3 & 1994-12-10 & 8400 & 5303\\
  \hline
\end{tabular}
\caption{\footnotesize Summary of {\it HST} observations of Abell 2390.}
\label{obs}
\end{table}

\begin{table}[htbp] \footnotesize
\begin{tabular}{l c c c c c}
\hline

Band & Image-plane & \multicolumn{2}{c}{Arc A, source-plane} & \multicolumn{2}{c}{Arc D, source-plane} \\
& & Major axis & Minor axis & Major axis & Minor axis\\
\hline
$V_{555}$ & 0.14\arcsec & 0.08\arcsec & 0.02\arcsec & 0.10\arcsec & 0.04\arcsec \\
$I_{814}$ & 0.17\arcsec & 0.10\arcsec & 0.03\arcsec & 0.12\arcsec & 0.06\arcsec\\
$J_{1250}$ & 0.21\arcsec & 0.12\arcsec & 0.04\arcsec & 0.15\arcsec & 0.07\arcsec\\
F126N & 0.22\arcsec & 0.12\arcsec & 0.04\arcsec & 0.15\arcsec & 0.07\arcsec \\
\hline
\end{tabular}
\caption{\footnotesize FWHM of the PSF in the image- and source-plane for each band.  For Arc A, source-plane values are given near the center of the image.  The $J_{1250}$-band and F126N PSFs are similar to within $<1\%$, in keeping with their near-identical pivot wavelengths.}
\label{psftab}
\end{table}

\begin{table}[htbp] \footnotesize
\begin{tabular}{l c c c c c c}
  \hline
  & Area only & H$\alpha$ & $J_{1250}$ & $I_{814}$ & $V_{555}$ & \citet{Swinbank06} \\ 
  \hline
  Arc A & 9.68 & 10.1 & 11.1 & 11.7 & 11.9 & $12.6^{+0.6}_{-0.8}$\\
  Arc D & 4.19 & 4.19 & 4.20 & 4.20 & 4.19 & $6.7^{+0.4}_{-0.2}$\\
  \hline
\end{tabular}
\caption{\footnotesize Differential magnifications of arcs A and D.  The first column shows the ratio of the image-plane to source-plane area along, while the other columns show the flux-weighted magnification in each band.  For reference, we give the magnifications used by \citet{Swinbank06}.}
\label{mags}
\end{table}

\begin{table}[htbp] \footnotesize
\begin{tabular}{l c c c c c}
  \hline
  & $F_\nu$ (mJy) & $F_\nu$ & $F_\nu$ & $F_\nu$ & $F_\nu$\\
  Observed $\lambda$ & 100~$\mu$m & 160~$\mu$m & 250~$\mu$m & 350~$\mu$m & 500~$\mu$m \\
  \hline
  Arc A & 7.1$\pm$1.6(stat.)$\pm$0.33(confusion) & 10.3$\pm$3.1$\pm$1.47 & $<$18.2 & $<$14.9 & $<$24.8 \\
  Arc D & 21.7$\pm$1.6$\pm$0.33 & 35.3$\pm$3.1$\pm$1.47 & 24.7$\pm$1.7$\pm$7.3 & 24.0$\pm$1.6$\pm$8.6 & $<$22.7 \\
  \hline
\end{tabular}
\caption{\footnotesize Far infrared observed flux densities of the two arcs from the \emph{Herschel} PACS/SPIRE archival data \citep{Lutz11,Oliver12}.  Expected 1-$\sigma$ instrumental errors from HSPOT are quoted.
3-$\sigma$ upper limits are derived from aperture photometry for non-detections, and the expected confusion noise from HSPOT is also given for the values that are not upper limits.}
\label{tab: herschel phot}\end{table}

\begin{table}[htdp] \footnotesize
\begin{tabular}{l | c | c} 
\hline
& \bf{Arc A} & \bf{Arc D} \\
\bf{Method} & Log($L$(TIR)/L$_\odot$) & Log($L$(TIR)/L$_\odot$) \\
\hline
\citet{Rieke09} Table 4 & 10.62$\pm$0.05 & 11.42$\pm$0.02\\
Integrate 8-1000~$\mu$m & 10.62$\pm$0.05 & 11.40$\pm$0.02\\
$f$(24$\mu$m) + Equation 5 of \citet{Rujopakarn11b} & 10.81$\pm$0.08 & 11.39$\pm$0.05\\
Integrate 3-1100~$\mu$m & 10.69$\pm$0.05 & 11.42$\pm$0.02 \\
\hline
\end{tabular}
\caption{\footnotesize Comparison of methods to measure the total far infrared luminosity, $L$(TIR).  The values quoted in \citet{Rieke09} Table 4 use the method of \citet{Sanders03} to calculate $L$(TIR).  Unless otherwise stated, the far infrared luminosity spans 8-1000~$\mu$m.  All values are corrected for magnification, adopting the $\mu$ for H$\alpha$ given in Table \ref{mags}.}
\label{LTIR}
\end{table}%

\begin{table}[bhtp] \footnotesize
\begin{tabular}{ l | c c | c c}
\hline
& \multicolumn{2}{c}{\bf{Arc A}} & \multicolumn{2}{| c}{\bf{Arc D}}\\
 & SFR & A(H$\alpha$) & SFR & A(H$\alpha$) \\
\bf{Method} & M$_\odot$ yr$^{-1}$ & & M$_\odot$ yr$^{-1}$ \\ 
\hline
L(8-1000~$\mu$m) \citep{Kennicutt98} + \citet{Rieke09} & $3.7^{+0.5}_{-0.4}$ & $1.3\pm0.2$ & $29\pm1$ & $1.8\pm0.1$ \\ 
$f(24 \mu$m) \citep{Rujopakarn11b} & 6$\pm$1 & 1.8$\pm$0.2 & $24\pm3$ & $1.6\pm0.1$ \\
L(3-1100~$\mu$m) + H$\alpha$ \citep{Kennicutt09} & $3.5\pm0.7$ & $1.2\pm0.2$ & 19$\pm$3 & 1.3$\pm$0.2 \\ 
\hline
\end{tabular}
\caption{\footnotesize Far-infrared star formation rates and the corresponding values of A(H$\alpha$), using each of the three methods described in \S \ref{FIR}: the \citet{Kennicutt98} formula with a correction for cold dust heating based on \citet{Rieke09} Figure 8; the observed 24~$\mu$m flux with the \citet{Rujopakarn11b} formalism; and the \citet{Kennicutt09} formalism to combine a fraction of the observed far-infrared luminosity with the extincted H$\alpha$ luminosity.}
\label{IRSFR}
\end{table}

\begin{table}[bhtp] \footnotesize
\begin{tabular}{ l | c c | c c}
\hline
& \multicolumn{2}{c}{\bf{Arc A}} & \multicolumn{2}{| c}{\bf{Arc D}}\\
& L & Extincted SFR & L & Extincted SFR  \\
\bf{SFR indicator} & (erg s$^{-1}$) & (M$_\odot$ yr$^{-1}$) & (erg s$^{-1}$) & (M$_\odot$ yr$^{-1}$) \\
\hline
H$\alpha$ & $(0.22 \pm 0.02)\times 10^{42}$ & $1.11 \pm 0.14$ & $(1.05 \pm 0.07)\times 10^{42}$ & $5.45 \pm 0.34$ \\ 
UV (2800 \AA) & $(0.78 \pm 0.06)\times 10^{45}$ & $0.67 \pm 0.05$ & $(2.58 \pm 0.06)\times 10^{45}$ & $2.21 \pm 0.05$ \\ 
\hline
\end{tabular}
\caption{\footnotesize Rest-H$\alpha$ and rest-UV luminosities and extincted (observed) SFRs for each arc.  All quantities have been de-magnified using the values from Table \ref{mags}.  The UV luminosity, quoted at 2800 \AA, has been multiplied by (1+z) to correct for bandwidth compression.}
\label{hauvsfrtab}
\end{table}

\begin{table}[htbp] \footnotesize
\begin{tabular}{l l c c }
  Rest-H$\alpha$ \\
  \hline
  & Ap. corr. & Flux & SFR \\
  & & $10^{-18}\ \mathrm{erg\ cm^{-2}\ s^{-1}}$ & \Mpyr \\  
  \hline
  Blue clump 1 & 0.79 & $7.6\pm0.8$ & $0.16\pm0.02$ \\ 
  Blue clump 2 & 0.75 & $5.7\pm0.6$ & $0.12\pm0.01$ \\
  Blue clump 3 & 0.58 & $1.4\pm0.2$ & $0.03\pm0.01$ \\
  Blue clump 4 & 0.78 & $1.4\pm0.1$ &  $0.03\pm0.01$ \\
  Red clump & 0.71 & $4.7\pm0.6$ & $0.10\pm0.01$ \\
  \hline
  Total for Arc A &  & $52\pm 2$ & $1.11\pm0.14$ \\
  Diffuse component &  & $31\pm2$ & $0.67\pm 0.14$ \\
  \hline
  \hline\\
  Rest-2800 \AA\\ 
  \hline
  & Ap. corr. & Flux density & SFR \\
  & & $10^{-20}\ \mathrm{erg\ cm^{-2}\ s^{-1}\ \AA^{-1}}$ & \Mpyr \\  
  \hline
  Blue clump 1 & 0.74 & $7.3\pm1.2$ &$0.07\pm0.01$\\
  Blue clump 2 & 0.71& $9.9\pm1.0$ &$0.10\pm0.01$ \\
  Blue clump 3 & 0.58 & $2.9\pm0.6$ &$0.03\pm0.01$\\
  Blue clump 4 & 0.73 & $3.0\pm0.5$ &$0.03\pm0.01$\\
  Red clump & - - - & $\leq 4.5$ & $\leq 0.04$ \\
  \hline
 Total for Arc A &  & $69\pm5$ & $0.67\pm0.05$ \\
  Diffuse component & & $41-46$ & $0.40-0.44$ \\
  \hline
  \hline
\end{tabular}
\caption{\footnotesize Summary of star formation rates and extinction from UV and H$\alpha$ photometry for the clumps in Arc A defined in Figure \ref{arcAclumps}.  Flux and flux density are quoted post-aperture correction on the source-plane, where the 2800 \AA\ flux density in the red clump is a 3$\sigma$ upper limit.  The aperture correction shown here gives the fraction of the total flux enclosed within each region, assuming each clump is unresolved.  UV flux densities are corrected for bandwidth compression as in Table \ref{hauvsfrtab}.} 
\label{clumpphot}
\end{table}

\clearpage

\end{document}